\documentclass{aa}  
\include{journals}
\usepackage{graphicx}
\usepackage{subfigure}
\usepackage{natbib}
\bibpunct{(}{)}{;}{a}{}{,} 
\usepackage{txfonts}
\newcommand{\xmm}{XMM-{\it Newton}}
\newcommand{\ace}{ACE}
\newcommand{\suzaku}{{\it Suzaku}}
\newcommand{\chandra}{{\it Chandra}}
\newcommand{\rosat}{ROSAT}
\newcommand{\rxte}{{\it RXTE}}

\newcommand{\papone}{{Paper I}}
\newcommand{\cf}{\ion{C}{v}}
\newcommand{\cs}{\ion{C}{vi}}
\newcommand{\mage}{\ion{Mg}{xi}}
\newcommand{\net}{\ion{Ne}{x}}

\newcommand{\oxys}{\ion{O}{vii}}
\newcommand{\oxye}{\ion{O}{viii}}
\newcommand{\sif}{\ion{Si}{xiv}}
\newcommand{\nsix}{\ion{N}{vi}}
\newcommand{\ns}{\ion{N}{vii}}
\newcommand{\modtwo}{{Model-2}}

\mathchardef\mhyphen="2D

\def\redc{$\chi^{2}_{\mu}$}
\def\cratio{$R_{\chi}$}
\def\er{$\mathrm{R}_{E}$}
\def\fluns{$\mathrm{keV}\,\mathrm{cm}^{-2}\,\mathrm{s}^{-1}\,\mathrm{sr}^{-1}$}
\def\nObs{3012}

\def\nthisswcx{103}
\def\sumswcx{64}
\def\winswcx{39}
\def\maxrevn{1773}
\def\pcswcx{3.4}
\def\pcoldswcx{6.5}
\def\oplussix{O$^{6+}$}
\def\opluss{O$^{7+}$}
\def\opluse{O$^{8+}$}
\begin{document}
\title{Identifying XMM-{\it Newton} observations affected by solar
  wind charge exchange - Part II\thanks{Based on observations with
    XMM-{\it Newton}, an ESA Science Mission with instruments and
    contributions directly funded by ESA Member States and the USA
    (NASA).}}

   \subtitle{}

   \author{J.A. Carter\inst{1}
          \and
          S. Sembay\inst{1}
          \and
          A.M. Read\inst{1}
          }

   \offprints{J.A. Carter}

   \institute{Department of Physics and Astronomy, University of Leicester, 
 Leicester, LE1 7RH, UK\\
              \email{jac48@star.le.ac.uk}\\
	      \email{sfs5@star.le.ac.uk}\\
	      \email{amr30@star.le.ac.uk}\\
                }

   \date{Received ; accepted }

 
  \abstract
  {}
  {We wished to analyse a sample of observations from the XMM-{\it Newton}
    Science Archive to search for evidence of exospheric solar wind charge
    exchange (SWCX) emission.}
  {We analysed \nObs\ observations up to and including revolution
    \maxrevn. The method employed extends from that of the previously
    published paper by these authors on this topic. We detect temporal
    variability in the diffuse X-ray background within a narrow
    low-energy band and contrast this to a continuum. The low-energy
    band was chosen to represent the key indicators of charge exchange
    emission and the continuum was expected to be free of SWCX.}
  {Approximately \pcswcx\,\% of observations studied are
    affected. We discuss our results with reference to the XMM-{\it Newton}
    mission. We further investigate remarkable cases by considering
    the state of the solar wind and the orientation of XMM-{\it Newton} at
    the time of these observations. We present a method to approximate
    the expected emission from observations, based on given solar wind
    parameters taken from an upstream solar wind monitor. We also
    compare the incidence of SWCX cases with solar activity.}
  {We present a comprehensive study of the majority of the suitable
    and publically available XMM-{\it Newton} Science Archive to date, with
    respect to the occurrence of SWCX enhancements. We present our
    SWCX-affected subset of this dataset. The mean exospheric-SWCX
    flux observed within this SWCX-affected subset was 15.4 \fluns\ in
    the energy band 0.25 to 2.5\,keV. Exospheric SWCX is
    preferentially detected when XMM-{\it Newton} observes through the
    subsolar region of the Earth's magnetosheath. The model developed
    to estimate the expected emission returns fluxes within a factor
    of a few of the observed values in the majority of cases, with a
    mean value at 83\,\%.}


   \titlerunning{}
   \authorrunning{J.A. Carter et al.}
   \maketitle
%

\section{Introduction}\label{secintro}

This paper follows that of \citet{carter2008} (hereafter \papone),
whereby a set of approximately 180 \xmm\ observations, taken between
revolutions 52 and 1104 (March 2000 until December 2005), were
analysed to search for cases of solar wind charge exchange emission
(SWCX) occurring within the Earth's magnetosheath or in near
interplanetary space. In \papone\ we searched for time-variable SWCX
signatures and found that approximately \pcoldswcx\,\% of the
observations studied were affected. One case of SWCX showed an
extremely rich emission line spectrum, which was attributed to a
passing Coronal Mass Ejection (CME) and has been discussed in detail
by \citet{carter2010}. This paper extends the work of \papone, to
provide a complete sample, covering \nObs\ suitable \xmm\ observations
downloaded from the \xmm\ Science Archive
(XSA)\footnote{http://xmm.esac.esa.int/xsa/}. We include data from
between revolutions 28 and \maxrevn\ (February 2000 until August
2009).

SWCX processes occur at many locations within the solar system
including planetary magnetosheaths, the corona of comets, within the
heliosphere and at the heliospheric boundary where the outer reaches
of the interplanetary magnetic field encounter that of the surrounding
interstellar medium. Those charge exchange processes that result in
the emission of X-rays occur for interactions involving a
highly-charged solar wind ion, for example the bare oxygen ion
\opluse, and a donor neutral species, such as hydrogen in the case of
geocoronal SWCX emission.

The possibilities for viewing exospheric SWCX emission in the vicinity
of the Earth depend on the orbital and viewing constraints of the
observatory in use, however one may be able to detect enhancements due
to charge exchange occurring within the heliosphere. This will show
fluctuations on longer timescales to SWCX occurring within the Earth's
exosphere \citep{cravens2001}. SWCX emission from within the
heliosheath (for example resulting from the helium-focusing cone) is a
contributor to the X-ray emission from the supposed Local Hot Bubble
in which the Sun resides \citep{koutroumpa2007, koutroumpa2008}, and
to what level it contributes is under great debate. We concentrate,
however, on SWCX emission occurring in the near vicinity of the Earth
and use the time variable nature of this emission as our marker for
selecting affected \xmm\ observations.

In addition to the pioneering work noting the so-called Long Term
Enhancements (subsequently known to be due to SWCX emission) within
\rosat\ observations \citep{snowden1995}, the assignment of such
enhancements to SWCX emission has also been observed in studies
undertaken using data from \suzaku\ \citep{fujimoto2007, bautz2009}
and \chandra\ \citep{smith2005}. Enhancements in soft X-ray band \xmm\
spectra have been attributed to SWCX contamination in the literature
\citep{snowden2004, kuntz2008, snowden2009a, henley2008}. This has
almost exclusively involved the comparison of multiple pointings of
the same field which has enabled the serendipitous detection of the
low-energy enhancement, most notably around the \oxys\ helium-like
triplet at approximately 0.56\,keV. Observations of the Groth-Westfall
Strip, Polaris Flare region and the \textit{Hubble} Deep Field as
discussed in \citet{kuntz2008} were included in the analysis of
\papone\ as control cases for our method.

The \xmm\ observatory \citep{jansen} has the largest collecting area
in the band 0.2 to 10\,keV of all X-ray telescopes currently in orbit
and the European Photon Imaging Camera (EPIC) suite of instruments on
board (two MOS \citep{turner} and one pn \citep{struder} CCD cameras)
provide medium spectral resolution ($\frac{E}{\Delta E} \sim$
17). During various periods of its orbit and depending on
observational constraints, \xmm\ may view regions of the Earth's
magnetosheath which are predicted to exhibit the highest X-ray
emissivity due to SWCX between highly charged solar wind ions and
hydrogen in the Earth's exosphere \citep[and references
therein]{robertson2006}. We use data from the EPIC-MOS cameras to
identify times of variability in a low energy band that is not
mirrored in a higher energy band, to select those periods we suspect
have been affected by variable SWCX.

SWCX emission can be considered as a contaminant by sections of the
community wishing to study Galactic or Extragalactic sources beyond
the heliosheath. Understanding the occurrence and nature of the
emission is important therefore for the process of elimination from
astronomical observations. However, SWCX emission can be used to
provide diagnostics of the solar wind. X-rays emitted from the comas
of comets have been suggested and used for this purpose
\citep{cravens1997, dennerl1997, lisse2001, bodewits2007}. A detailed
study of the heavy ion constituents of a CME was discussed in
\citet{carter2010}. In addition, exospheric-SWCX may be used to test
models of dynamical processes within the Earth's magnetosheath and the
solar-terrestrial connection, such as flux-transfer events or boundary
layer phenomena \citep{collier2010}.

The layout of the paper is as follows. In Section~\ref{secmethod} we
describe the method employed to identify those observations of
interest. In Section~\ref{secresgen} we present the overall results
for the whole sample studied and compare the occurrence of SWCX with
the solar cycle. In Section~\ref{secspec} we describe fitting spectral
models to cases of SWCX enhancement data. In Section~\ref{seccases} we
discuss in more detail several of those observations affected by
SWCX. In Section~\ref{secemiss} we describe a model of the expected
X-ray emission using the orbit of \xmm\ and the solar wind conditions
at the time of the observation. We finish with our discussion and
conclusions in Section~\ref{secconcl}.

\section{Data analysis}\label{secmethod}

All observational data used in this study are available through the
XSA where we extract the Original Data Files (ODF). We consider
observations taken by the EPIC-MOS \citep{turner} cameras (MOS1 and
MOS2) using the full-frame mode only. Therefore the observations used
provided an even sample across the mission and the event list data
from the cameras could be combined if the same filter was used for
both. We considered observations up to and including revolution
\maxrevn\ (August 2009).

Individual observation event lists were created from the Original Data
Files for each EPIC-MOS camera and filtered for soft-proton
contamination using the publicly available Extended Source Analysis
Software
(ESAS)\footnote{http://heasarc.gsfc.nasa.gov/docs/xmm/xmmhp\_xmmesas.html}
package (\textit{mos-filter} tool). At the time of data processing,
ESAS was only available for the EPIC-MOS cameras. This tool fits a
Gaussian to a histogram of in-field-of-view count rates in a high
energy band (2.5\,keV to 12.0\,keV). Time periods with count rates
beyond a threshold of ${\pm}$1.5\,$\sigma$ away from the peak of this
Gaussian were removed by applying a Good Time Interval (GTI) file to
the event lists. The GTI files created for each EPIC-MOS event list
were combined together to form one EPIC-MOS GTI file and this file was
reapplied to both the MOS1 and MOS2 event lists to provide
simultaneous coverage during each observation. Resolved point sources
(from lists created for the 2XMM catalogue \citep{watson2009} using a
minimum likelihood threshold $\geq$ 6), were removed from the field of
view by extracting events in a circular region of 35 arcseconds radius
about the source position. Those observations judged, after a visual
inspection (and prior to creation of any spectra, see
Section~\ref{secspec}), to show residual source contamination (the
wings of the point spread function of the EPIC-MOS camera being
evident in an image of the event file) passed through an additional
spatial filtering stage using a larger extraction region to further
clean the dataset. A more detailed description of the filtering steps
and nature of soft proton contamination can be found in \papone\ and
\citet{carter2010}.

\begin{figure*}
  \centering
  \includegraphics[width=0.47\textwidth, bb=54 362 513 688]{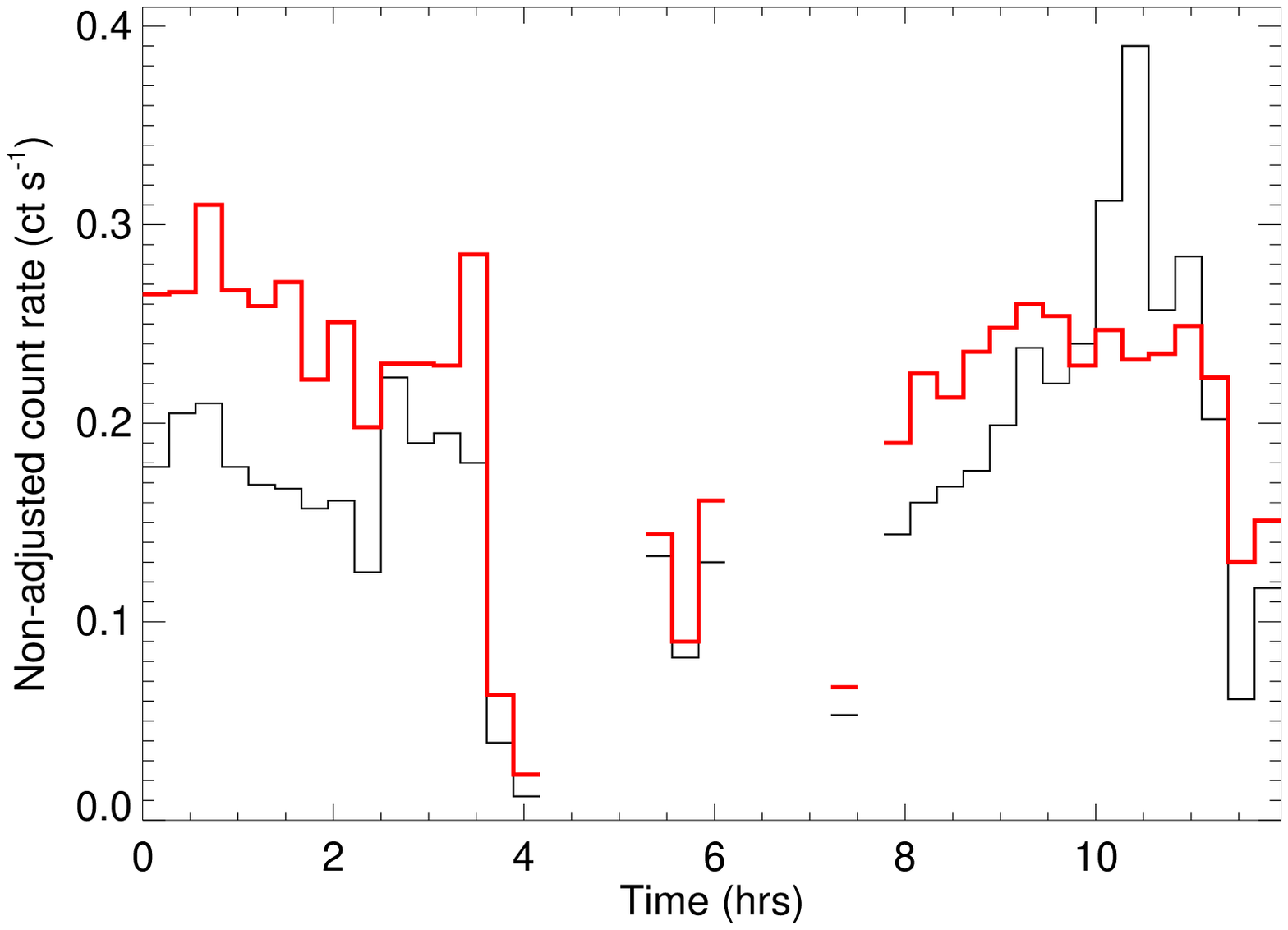}
  \includegraphics[width=0.47\textwidth, bb=54 362 513 688]{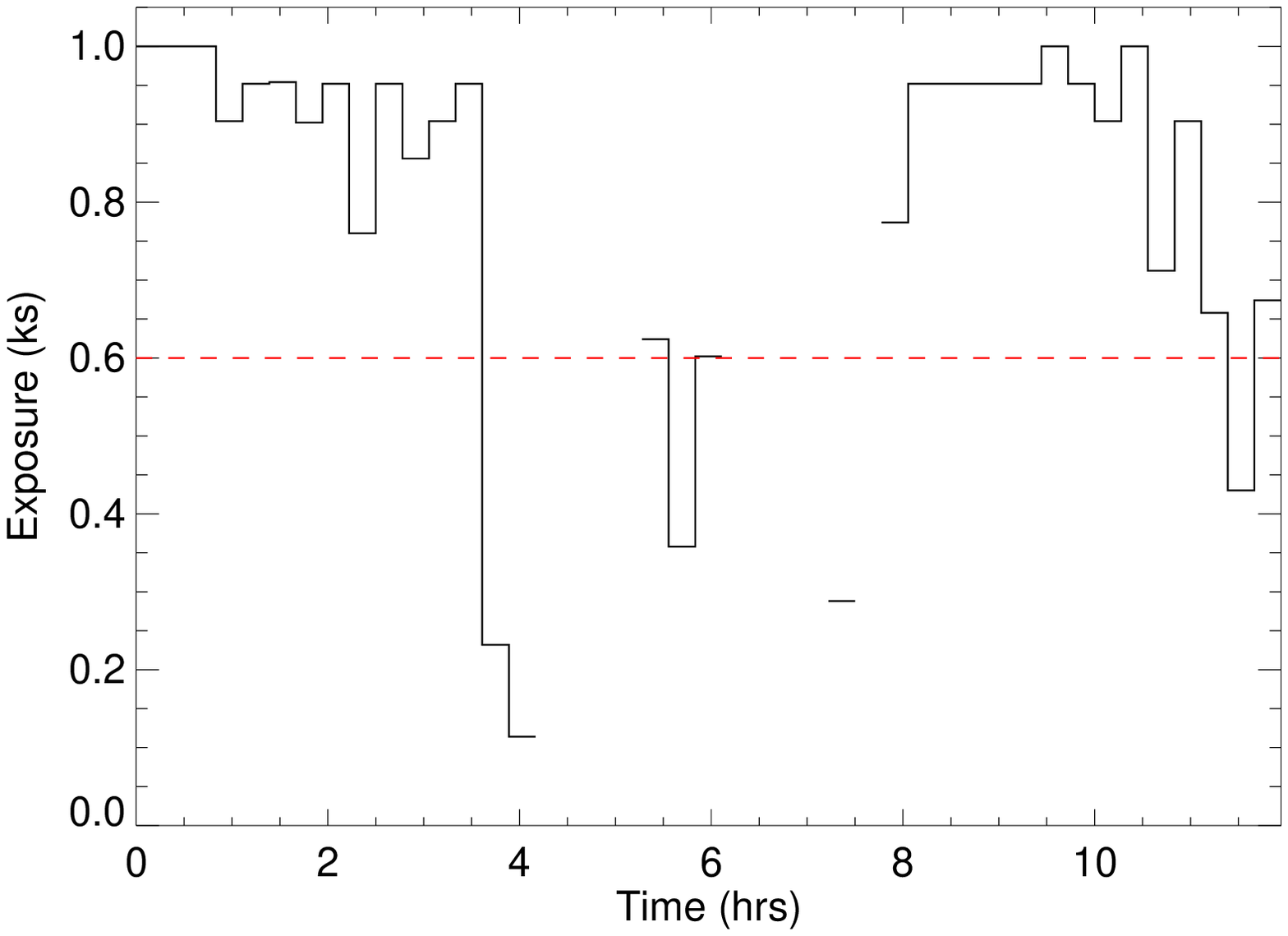}
  \includegraphics[width=0.47\textwidth, bb=54 362 513 688]{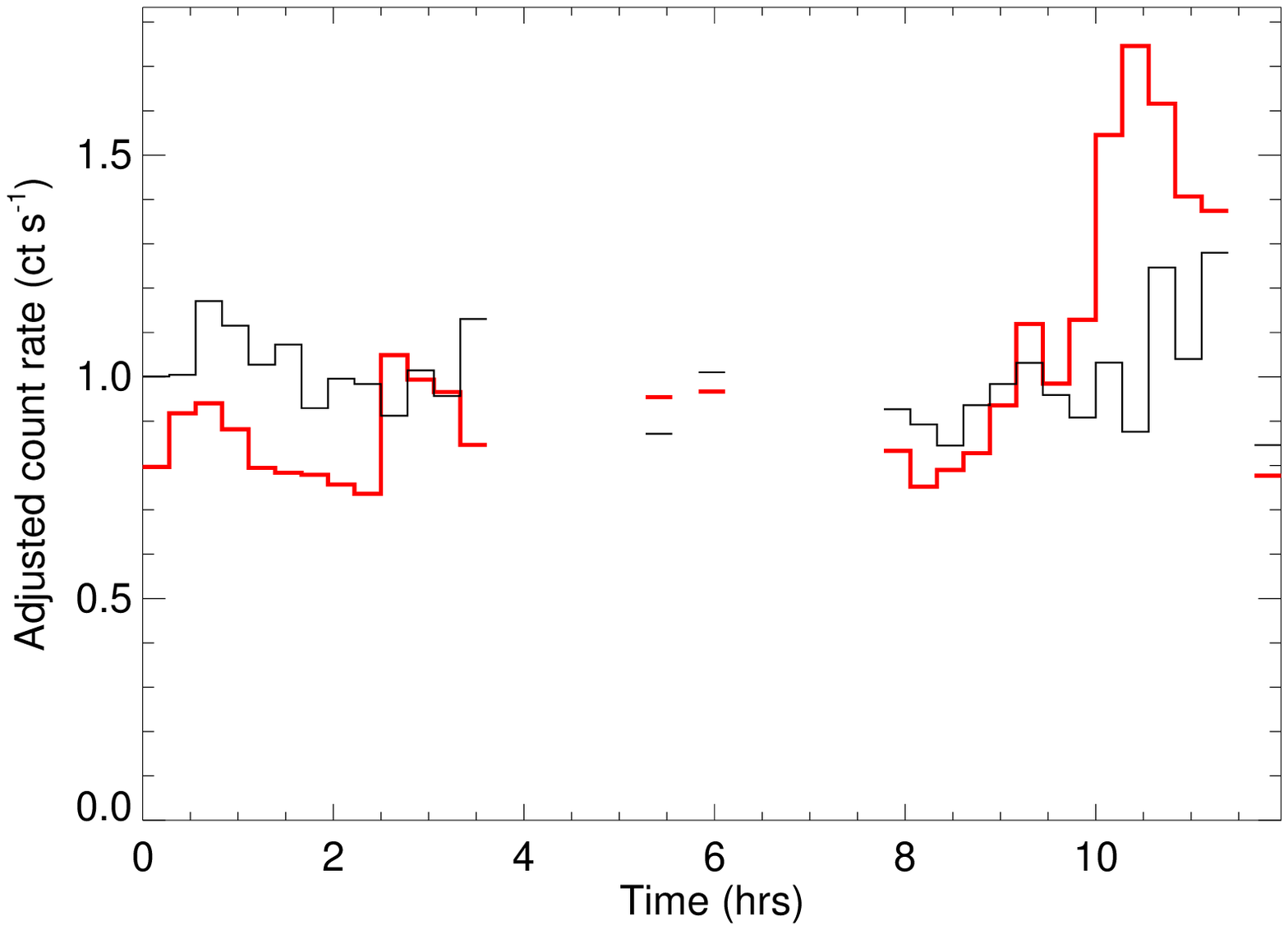}
  \includegraphics[width=0.47\textwidth, bb=54 362 513 688]{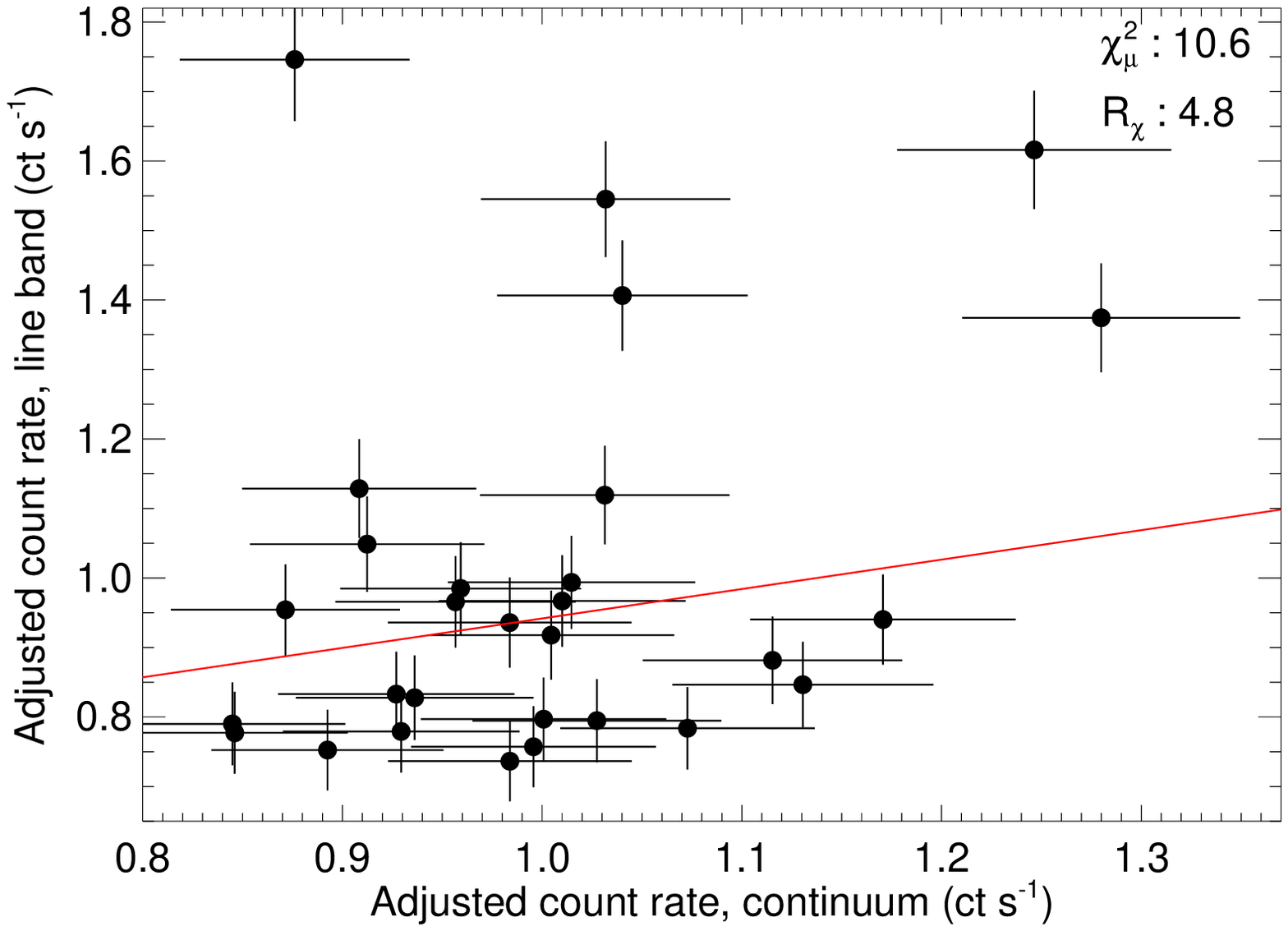}
  \caption{Lightcurve correction procedure example for observation
    with identifier 0150680101 (line-band (black), continuum (red) for
    panels top-left and bottom-left). Top left: example lightcurves
    showing a peak in the line-band that is not reflected in the
    continuum. Top right: exposure coverage for each bin, the
    threshold at 60\% is marked by the red dashed line. Bottom left:
    lightcurves after the adjustments for exposure correction and
    scaling by the mean. Bottom right: example scatter plot for this
    observation.}
  \label{figexamlc}
\end{figure*}

The method employed here followed the first steps as described in
detail within \papone. In summary, two lightcurves with bin size of
1\,ks were created for each observation from events within the full
field-of-view (radius of 13.3 arcminutes). When both EPIC-MOS cameras
were used during an observation and employed the same filter, events
from both cameras were used to construct the lightcurves. The first
lightcurve was chosen to represent the continuum, covering events with
energies in the range 2.5\,keV to 5.0\,keV. The second lightcurve was
extracted using events with energies in the range 0.5 to 0.7\,keV to
cover the strong SWCX emission from \oxys\ and \oxye\ (the line-band
lightcurve). This energy range incorporates emission energies from the
\oxys\ triplet and resonance lines which are dominated by the
forbidden line transition at 0.56\,keV. Lightcurves were then
exposure-corrected for periods removed during the filtering steps. As
the MOS1 and MOS2 event files for each observation, prior to the
lightcurve creation, were filtered using a single GTI file which is
not energy specific, the exposure-coverage for each bin was the same
for the line-band and continuum lightcurves. Therefore the same
exposure-correction factor for an individual bin was applied to both
lightcurves in this step. We keep bins of the lightcurve with at least
60\% of the full exposure for that bin and reject the remaining bins
(in contrast to \papone\ when we kept all bins with at least 40\%
exposure). Lightcurves were rejected from further analysis if they
were less than 5\,ks in length. By increasing the strictness of the
bin coverage thresholds, we reduced the incidence of type I errors
(those incorrectly labelled detections), but ran the risk of
increasing the number of cases that were incorrectly labelled as
non-detections (i.e. not showing a deviance from the null hypothesis
of a linear fit between the line-band and continuum (type II
errors). We scaled each lightcurve by its mean to produce adjusted
lightcurves. The count rates for the line-band and the continuum band
were then always of the same order which facilitated the
identification of periods of enhancement in the line-band
lightcurve. An example of the combined-MOS lightcurve adjustment
process can be found in Figure~\ref{figexamlc} (panels top-left,
top-right and bottom-left).

We plotted a scatter plot between the two bands (using the adjusted
line-band as the dependent variable), shown in Figure~\ref{figexamlc}
(bottom-right). A linear model fit to each scatter plot was computed
using the IDL procedure, \textit{linfit}, which minimises the
$\chi^{2}$ statistic.

We computed the reduced-$\chi^{2}$ for the fit, hereafter referred to
as \redc, by dividing the $\chi^{2}$ by the number of bins minus one,
to account for the reduction in the number of degrees of freedom made
by fitting a linear model to the data. A high \redc\ indicates that a
significant fraction of the points deviate significantly from the best
fit line. We expected these cases would be more likely to show
variable SWCX-enhancement. In addition we computed the $\chi^{2}$
values for each individual lightcurve in terms of the deviation from
the mean of that lightcurve. We calculated the ratio between the
line-band and continuum $\chi^{2}$ values to add to our diagnostic
(hereafter denoted as \cratio).

\nObs\ observations made up the final sample to be used for further
analysis. The results from \papone\ showed us that those observations
exhibiting both high \redc\ and high \cratio\ were most likely to show
near-Earth time-variable SWCX signatures. Observations that fulfilled
these criteria were considered for further analysis, spectrally,
temporally and with regard to the orientation of \xmm.

\section{Global results}\label{secresgen}

All observations in our final sample, after any rejections as
described below, were ranked by \redc. The two highest ranked
observations were observations of comets. Although resolved point
sources have been removed, cometary X-rays are diffuse and will likely
be spread over a large fraction, if not all of the field of view. We
assume that the dominant variations in the line-band lightcurve that
result in such high ranking are due to SWCX emission occurring within
the cometary coma and not to any emission occurring within the
vicinity of the Earth. We discuss the cometary cases in
Section~\ref{seccomets}.

After ranking the observations, we were able to study those that
exhibited the highest \redc\ and \cratio\ in more detail on a case by
case basis. 

Observations were examined for residual point or extended sources that
may contribute to the high variations in the line-band. Extended or
diffuse residual sources that remain in the field of view will not
affect this detection method providing no inherent variation occurs
within these sources, as expected for sources outside the solar
system, in either one of the bands. Cases were also examined for
residual soft proton contamination. Although the files used in this
analysis have been filtered for periods of soft proton flaring,
residual contamination may remain. Excessive scatter due to residual
soft proton contamination impedes the ability to identify periods of
exospheric SWCX and will result in some type II errors in our
sample. Also, if exospheric SWCX occurs throughout the entirety of the
observation little or no variation can be seen in the line band,
resulting in a rejection of this case. Excessive and simultaneous
variations in both the line-band and continuum will result in a high
\redc\ yet a low value of \cratio. We concentrate our analysis
therefore on cases that exhibit both high \redc\ and high \cratio.

For some observations with short lightcurves, it was impossible to
identify any time-periods of boosted line-band emission (the putative
SWCX enhancement periods). These observations were disregarded as
SWCX-enhancement cases. Also, the method described in this paper
tested only for variable SWCX on short timescales. Spectral analysis
of suspected SWCX cases was therefore only possible when a clear
line-band enhancement period could be identified during the duration
of the observation.

We find \nthisswcx\ observations in our sample that show indications
of a time-variable exospheric SWCX enhancement, that are not excluded
from consideration based on the reasons described above. All of these
cases had a \redc\ value greater than or equal to 1.2 and a \cratio\
value of greater than or equal to 1.0. These cases make up only
$\sim$20\% of all observations that have values of \redc\ and \cratio\
above these thresholds, indicating that although the values of \redc\
and \cratio\ are indicators of a SWCX-enhancement, considerable
inspection of an observation on a case-by-case basis is still
required. The majority of cases in the whole sample have a \cratio\
value less than 1, indicating that there is more variation seen in the
continuum compared to the line-band. The continuum incorporates the
break energy at $\sim$3.2\,keV in the two-power law models of residual
soft proton contamination \citep{kuntz2008}. For higher intensity
soft-proton flares, the slope of the power-law becomes
flatter. Therefore the higher energy and wider continuum will have a
greater variance than the softer, narrower line-band due to the
presence of unfiltered residual soft protons.

The top 10 observations as ranked by \redc\ are given in
Table~\ref{tabtop} (excluding those observations that had been
rejected from consideration). Three of these top ten result from
cometary observations, four were identified in the work presented in
\papone\ and three are new observations discovered during this
analysis. Many of the highest ranked cases had previously been
identified in the literature. It is our intention to make the full
ranked list of observations used in this analysis publically available
online in the future, most probably through the \xmm\ EPIC Background
Working Group (BGWG)
\footnote{http://xmm.vilspa.esa.es/external/xmm\_sw\_cal/background}. We
include a summary table listing the SWCX cases in
Table~\ref{tabappendix}.

\begin{table}
  \caption[]{Highest ranked observation by \redc. Those observations with severe residual soft proton contamination have been excluded.} 
\begin{tabular}{llll}
  \hline\hline
  Revn. & Obsn.      & \redc\ & Comment  \\ 
  \hline
  0369  & 0103461101 & 856.7 & Comet C2000 WM1 (LINEAR)   \\
  0808  & 0164960101 & 226.6 & Comet C2001 Q4 (Neat)  \\
  0342  & 0085150301 &  27.2 & \papone\ \& \citet{carter2010}\\
  0209  & 0093552701 &  23.0 & \papone \\
  1014  & 0305920601 &  15.0 & \papone \\
  0690  & 0149630301 &	14.1 & \papone \\
  0623  & 0150610101 &  13.5 & New case \\
  1177  & 0406950201 &  13.3 & Comet 73p \\
  0339  & 0054540501 &  13.2 & New case \\	
  0422  & 0113050401 &	12.7 & New case \\
\end{tabular}
\label{tabtop}
\end{table}

\begin{figure}
  \centering
  \includegraphics[width=0.475\textwidth, angle=0, bb=56 360 585 740]{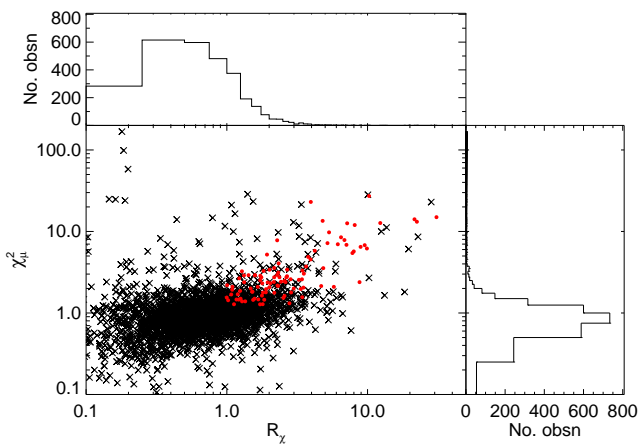}
  \caption{\redc\ versus \cratio. Red filled circles indicate those
    observations with time-variable SWCX signatures. Histograms of the
    \redc\ and \cratio\ values are shown in the side panels (using a
    bin size of 0.25).}
  \label{figresscat}
\end{figure}

A scatter plot of \redc\ versus \cratio\ is given in
Figure~\ref{figresscat}. Those observations exhibiting both high
\redc\ and high \cratio\ but which do not show clear SWCX signatures
(i.e. an enhancement in the mean-adjusted line-band lightcurve
compared to that of the continuum) have been rejected from further
analysis. We discuss the detection methods of SWCX cases in the
literature, other than by a search for a time-variable low-energy
component in Section~\ref{secmulti}. Cometary X-ray emission is
discussed in Section~\ref{seccomets}.

\begin{figure}
  \centering
  \includegraphics[width=0.475\textwidth, bb=79 370 565 700]{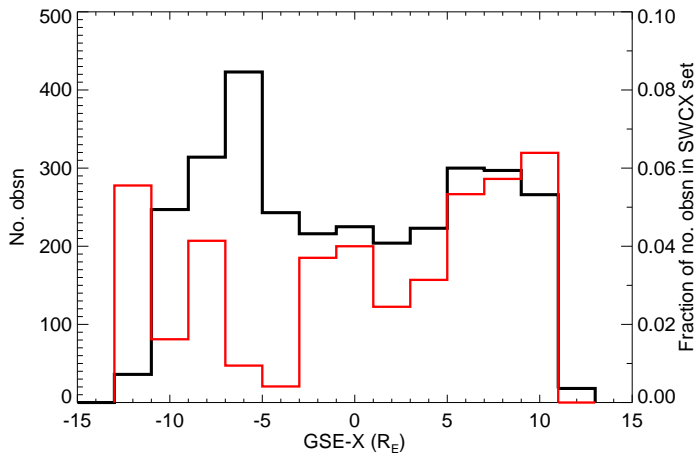}
  \caption{Total number of observations (black) versus GSE$\mhyphen$X
    position and the fraction of observations detected with exospheric
    SWCX enhancements (red).}
  \label{figgsexhistfrac}
\end{figure}

In Figure~\ref{figgsexhistfrac} we plot the total number of
observations and the fraction of observations that show SWCX
enhancements, versus the GSE$\mhyphen$X position of \xmm\ at the
mid-point of each observation. We can see from this figure that \xmm\
is preferentially found on the subsolar side of the Earth when SWCX
enhancements occur. In addition, the largest bin of the fractional
plot occurs around the nominal magnetopause stand off distance, at
approximately 10\,\er. The \xmm\ line-of-sight for these cases
traversed the subsolar region (sunward side) of the magnetosheath, as
expected according to the modelling work of
\citet{robertson2006}. \xmm\ viewing is constrained by the fixed solar
panels and limits imposed to avoid directly observing the Sun, Moon
and Earth and is only able to view the sunward side of the
magnetosheath at certain times of the year \citep{carter2008}.

We also consider the seasonal variation of the occurrence of SWCX in
this sample. \sumswcx\ SWCX cases occurred during the summer months
(April until September inclusive) and \winswcx\ during the winter
(October until March). More exospheric SWCX cases are expected in the
summer months, as viewed by \xmm\ \citep{carter2008}.

\subsection{Relationship with the solar cycle and solar  wind}\label{secsolar}

\begin{figure}
  \includegraphics[width=.48\textwidth, angle=0, bb=60 362 575 722]{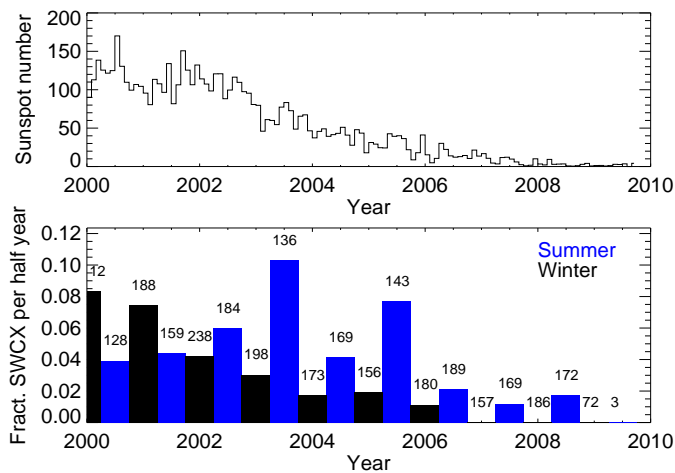}
  \caption{Top panel: sunspot number versus time. Bottom panel: the
    coloured histogram of the fraction of observations affected by
    exospheric SWCX is binned into six month periods (blue - summer,
    black - winter). The total number of all observations for each
    period is noted above the bin.}
  \label{figsolarcycle}
\end{figure}

In Figure~\ref{figsolarcycle} we plot the solar sunspot
number\footnote{http://www.sidc.be/index.php} from the latter half of
Solar Cycle 23 and mark the times at which \xmm\ observations with
known SWCX (this paper, paper I and \xmm\ exospheric-SWCX cases in the
literature) occurred. We plot a histogram of the number of SWCX cases
per half year, to remove any bias resulting from the seasonal
constraints on pointing angle experienced by \xmm. Each histogram bin
starts at the start of summer ($1^{st}$ April) or start of winter
($1^{st}$ October). As expected for cases of exospheric SWCX, there
are more cases at times of high solar activity around solar maximum
than when approaching solar minimum. Also, for the cases from 2002
onwards and approaching solar minimum, we see a higher proportion of
SWCX cases in the summer six-month period compared to the winter
period, for the same year. The significance of this trend however
should not be overstated due to the low number of cases in each bin.

\subsection{Multiple pointings of target fields}\label{secmulti}

Multiple pointings towards the same target allow one to compare
diffuse and extended emission spectral models that may exhibit
spectral variations indicative of SWCX contamination. The long term
enhancements of the ROSAT all-sky maps, which were subsequently
attributed to SWCX, were first identified by comparisons between
fields \citep{snowden1995}. \citet{kuntz2008} examined multiple \xmm\
observations of the \textit{Hubble} Deep Field, amongst other targets,
and identified a proportion of their set affected to different degrees
by SWCX emission. \citet{bautz2009} inferred SWCX-enhancements in
observations by \suzaku\ towards the cluster Abell 1795 after
examination of a low-energy lightcurve revealed peaks coincident in
time with enhancements in the solar wind proton flux as measured by
\ace. \citet{henley2010} used a large set of \xmm\ observations to
compare intensities from \oxys\ and \oxye\ lines from sets of
observations with the same target pointings. They find no universal
association between enhanced SWCX emission and the closeness of the
the line-of-sight to the sub-solar region of the magnetosheath. In
this paper we do see a tendency for \xmm\ to be clustered around the
sub-solar region for the SWCX cases, as discussed in
Section~\ref{secresgen}.

One of the \citet{henley2010} SWCX-enhanced cases is part of the data
set used in this paper, however this was not detected by our method as
no discernable variability occurred during the observation. Cases of
SWCX occurring in the Earth's exosphere, identified by detecting
time-variable emission, have also been observed by \suzaku\
\citep{fujimoto2007, ezoe2010}. \citet{carter2010} used a previous
observation of a target field to constrain the diffuse X-ray emission
inherent to that look direction to calculate the strength of SWCX
emission lines associated with a CME passing in the vicinity of the
Earth. The data set presented in the present paper contains a number
of previously known cases of SWCX, however, not all of these were time
variable and therefore were not picked up by the technique used
here. They did however become tests of the ability of this method to
identify SWCX-affected observations. One limitation of this technique
is that SWCX emission occurring at an approximately steady state as
part of any quiescent geocoronal X-ray emission will not be
identified. In addition, any observed quiescent heliospheric SWCX will
in general be several times stronger than any quiescent geocoronal
X-ray emission, due to the increased integration lengths involved
\citep{cravens2001}. Although a combination of techniques would be
ideal, this is not possible for the vast majority of single \xmm\
pointings. The greatest advantage of this method is that observations
are considered on an individual basis. An \xmm\ user can make a
judgement as to whether extra caution is required when analysing their
results for possible SWCX-contamination.

\subsection{Cometary emission}\label{seccomets}

Comet Hyakutake was the first comet whose X-ray emission, as observed
by \rosat\ and \rxte\ \citep{lisse1996}, was assigned to the SWCX
emission process \citep{cravens1997}. This SWCX emission occurs from
the interaction of the solar wind with neutral species that outgas
from the comet as it enters the inner solar system, and the amount of
out-gassing is dependent on the comet's distance from the Sun. These
neutral species are mainly water and its dissociation products. The
SWCX emission must occur in cometary regions where photoionisation and
destruction of the neutral species can occur. Water and hydroxyl ions
have short lifetimes when exposed to solar UV photons and therefore
survive the longest in the coma interior, whereas the dissociation
products of water (along with CO providing the comet has a
sufficiently high carbon abundance) can survive further into the outer
coma regions. A detailed description of X-ray emission from comets,
primarily using data from the \chandra\ observatory, can be found in
\citet{bodewits2007, bodewits2007b}.

\begin{figure}
  \includegraphics[width=0.495\textwidth, angle=0, bb=84 380 544 706]{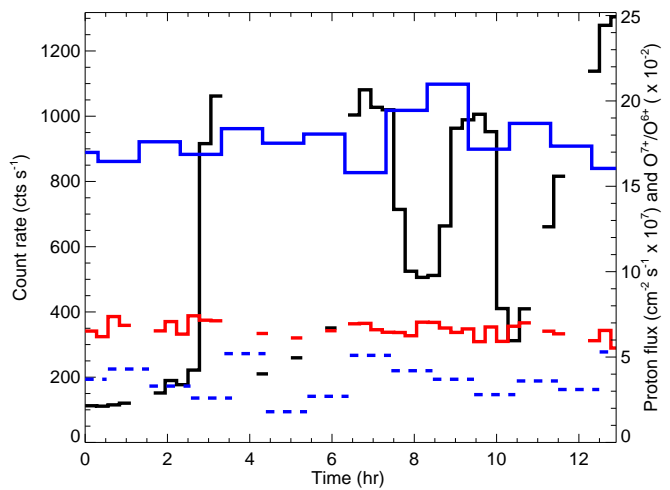}
  \caption{Lightcurve from comet C2001 Q4 (Neat) that resulted in the
    highest \redc\ values (black - SWCX band, red - continuum
    band). The time axis is given in hours since the start of the
    \xmm\ lightcurves. The solar wind proton flux as recorded by \ace\
    is given in blue. The \opluss\ to \oplussix\ ratio is plotted
    using the proton axis and is shown by the blue dashed line.}
  \label{figcomets}
\end{figure}

Several \xmm\ observations of comets were included in the sample in
this paper. The \redc\ and \cratio\ values for the comets are given in
Table~\ref{tabcomets}. The highest overall values of \redc\ and
\cratio\ occurred during observations of comets. Example line-band and
continuum lightcurves from comet C2001 Q4 (Neat) (observation
0164960101) are shown in Figure~\ref{figcomets}, where the line-band
lightcurve clearly dominates and is highly variable. Also, an earlier
observation of the same comet exhibited very different values of
\redc\ and \cratio. Although the upwind solar wind monitor Advanced
Composition Explorer (\ace\, level 2, combined instrument data,
\citet{stone1998}) proton flux, also shown in this graph, is steady
and not remarkable in intensity, the comet will more likely be
sampling solar wind that originates from a different location in the
solar corona. In Figure~\ref{figcomets} we have also plotted the
\opluss\ to \oplussix\ ratio, using data from the \ace\ SWICS
instrument. This ratio varies slightly over the length of the
observation but any fluctuations are not reflected in the \xmm\
line-band lightcurve, further implying that the comet is sampling a
different solar wind to that seen by \ace. The solar wind is a
collisionless plasma and so its ion composition remains unchanged as
it flows away from the Sun. Signatures of SWCX occurring throughout
the solar system can therefore be used to infer the composition of the
solar wind, which varies considerably throughout the solar cycle and
with solar latitude. As cometary orbits are not restricted to the
ecliptic plane they are ideal locations to study compositional
signatures from solar wind originating from a variety of solar wind
latitudes \citep{dennerl1997}. \citet{bodewits2007} was able to use
cometary X-rays to distinguish emission resulting from three solar
wind types: the cold and fast wind, the warm and slow wind and the
warm and disturbed wind. A more complete discussion of cometary X-ray
emission is beyond the scope of this paper.


\begin{table}
  \caption[]{\xmm\ observation of comets within the data set.} 
  \begin{tabular}{lllll}
    \hline\hline
    Revn. & Obsn.      & Name               & \redc   & \cratio  \\ 
    \hline
    0209  & 0103460901 & McNaught-Hartley   &   1.5  &   1.0 \\
    0369  & 0103461101 & C2000 WM1 (LINEAR) & 856.7  & 299.9 \\
    0719  & 0161760101 & Comet 2p (Encke)   &   2.0  &   1.5 \\
    0720  & 0113041301 & C2001 Q4 (Neat)    &   1.2  &   0.9 \\
    0808  & 0164960101 & C2001 Q4 (Neat)    & 226.6  & 349.77 \\
    1177  & 0406950201 & Comet 73p          &  13.3  &  17.2 \\
  \end{tabular} 
  \label{tabcomets}
\end{table}

\subsection{Planetary emission}\label{secplanets}
We include within the SWCX set an observation of the planet Saturn,
taken on $1^{st}$ October 2002. This observation has been
comprehensively analysed by \citet{branduardiraymont2010} who
attribute the X-ray flux to emission from the planetary disk, produced
by the scattering of solar X-rays and an additional fluorescent
emission line of oxygen at $\sim$0.53\,keV originating from the
rings. This observation was previously studied by \citet{ness2004} but
they did not investigate any low-energy variability. We find a high
\redc\ (2.9) and \cratio\ (2.3) in our time variability test and a
distinct step in the line-band lightcurve, indicative of a SWCX
enhancement, after the same filtering steps as to all other data sets
have been applied, including source exclusion to remove emission from
the planet and planetary exosphere. The initial source exclusion
radius equates to $\sim 3.7\,R_{Saturn}$. Spectra from this
observation were extracted from event files that had been additionally
filtered with a larger extraction region of $\sim
10.6\,R_{Saturn}$. Even after this additional source extraction step
the lightcurve production procedure yields metrics of \redc\ and
\cratio\ of 3.1 and 2.4 respectively. Two later observations of Saturn
taken in 2005 were included within our sample but showed no evidence
of SWCX enhancement. Therefore, we have no reason to reject this
observation from the SWCX set. We were concerned that the planet's
movement through and possibly out of the field of view may have caused
this effect. However, the dominant movement is in right ascension with
a maximum speed of 8\,arcseconds per hour
\citep{branduardiraymont2010}, resulting in a shift of only
0.78\,arcminutes over the course of the 21\,ks observation. In
addition, resolved sources have been removed from the field of view as
for all other observations.

\section{Spectral analysis}\label{secspec}

We continue our study by observing the spectral signatures of new SWCX
cases identified in this paper and in \papone. We omit those SWCX
observations that have been comprehensively investigated in the
literature yet show no temporal variability in our tests. We also omit
observations of comets. This set of observations which we have used
for further study throughout this paper is hereafter known as the SWCX
set and comprises \nthisswcx\ observations.

The Science Analysis System (SAS) software (version 9.0.0;
\textit{http://xmm.esac.esa.int/external/xmm\_data\_analysis/}) was
used to produce spectral products and instrument response files for
all observations of the SWCX set. The SAS accesses instrument
calibration data in so-called current calibration files (CCFs) which
are generally updated separately from SAS release versions. In this
paper we used the latest public CCFs released as of February 2010.

For each exospheric-SWCX case we extracted spectra for the EPIC-MOS
cameras for the suspected SWCX-affected period and for the suspected
SWCX-free period. The SWCX-affected period was when the enhancement in
the line-band lightcurve was judged (however not by a formal
mathematical argument) to have occurred. The enhancement could have
occurred at the beginning, middle or end of the observation. The
remaining time periods in the observation made up the SWCX-free
period. We used events from a circular extraction region, centred on
detector coordinate positions (DETX, DETY = -50, -180), with an
extraction radius of 16000 detector units or 13.3\,arcminutes. We also
applied the flag and pattern selection expression '\#XMMEA\_EM \&\&
PATTERN$<$=12 \&\& FLAG==0'. This pattern selection selects events
within the whole valid X-ray pattern library for the EPIC-MOS and the
flag selection removes events from or adjacent to noisy pixels and
known bright columns. We produced instrumental spectral response files
for each period. The instrument effective area files were calculated
assuming the source flux is extended, filling the field of view and
with no intrinsic spatial structure.
 
\subsection{Spectral modelling}\label{secspecmodel}

We knew from the work of \citet{carter2010} that exospheric SWCX can
occur throughout the entirety of an observation, although the
line-band lightcurve may show an enhanced and a steady-state
period. We used the spectra from the apparent SWCX-affected period as
the source spectra and that from the apparent SWCX-free period as the
background to produce a difference spectrum. The difference spectrum
therefore provides a lower limit to the SWCX enhancement that has
occurred during an observation. Providing the particle-induced
background is reasonably constant over the duration of the observation
(at most $\pm$10\% \citep{deluca2004}) this factor will be
eliminated. An inspection of the difference spectra was made for
energies above 2.5\,keV, to check for the presence of significant
variable residual soft-proton contamination. Each SWCX-case showed a
count rate statistically consistent with zero above this energy.

We modelled the resulting difference spectrum for each SWCX case with
a standard model of emission lines. The spectrum from MOS1 and MOS2
were fitted simultaneously, although a global normalisation parameter
for the MOS2 spectrum was allowed to vary. The relative line strengths
for a particular ion species below 1\,keV, for example \oxys\ (which
involves seven separate transitions including the \oxys\ triplet),
were set using the velocity dependent cross-sections of laboratory
charge exchange collisions between highly charged ion and atomic
hydrogen, as found in \citet{bodewits2007b}. We assumed a solar wind
speed of $400\,\mathrm{km}\,\mathrm{s}^{-1}$ for these
cross-sections. We also added emission lines from \net\ at 1.022\,keV,
\mage\ at 1.330\,keV and \sif\ at 2\,keV. There may be emission from
other ion species present in the spectra, such as from highly charged
iron or aluminium (as seen in \citet{carter2010}), but we wished to
simplify the model applied to a general case and the dominant SWCX
emission lines are found below 1\,keV. We fixed the relative
normalisations of the minor transitions to that of the principal
transition for each ion species. The principal, dominant transition in
the case of \cf, \nsix\ and \oxys\ is the forbidden line
transition. The principal ion transitions used in this modelling can
be seen in Table~\ref{tabmodel}.

\begin{table}
  \caption[]{Principle ion species emission lines used in the model, plus any minor emission line energies used (see text).} 
  \begin{tabular}{lll}
  \hline\hline
  Ion        & Energy (keV) & Minor energies (keV) \\ 
  \hline
   \cf       & 0.299        & 0.304, 0.308, 0.354, 0.379\\
   \cs       & 0.367        & 0.436, 0.459, 0.471\\
   \nsix     & 0.420        & 0.426, 0.431, 0.523\\
   \ns       & 0.500        & 0.593, 0.625, 0.640, 0.650\\
   \oxys     & 0.561        & 0.569, 0.574, 0.713, 0.666, 0.698, 0.723\\
   \oxye     & 0.653        & 0.775, 0.817, 0.837, 0.849\\
   \net      & 1.022        & \\
   \mage     & 1.330        & \\
   \sif      & 2.000        & \\
  \end{tabular}
  \label{tabmodel}
\end{table}

\begin{figure}
  \includegraphics[width=0.475\textwidth, angle=0, bb=66 360 535 656]{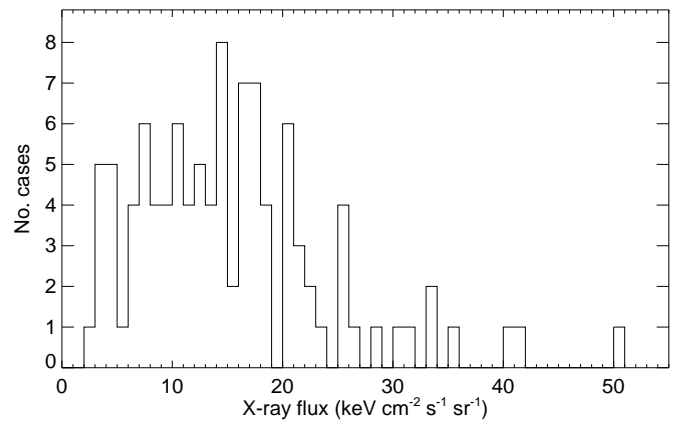}
  \caption{Histogram of total spectrally fitted flux between 0.25 and 2.5\,keV for the SWCX set.}
  \label{figfluxhist}
\end{figure}

We used Version 12.5.0 of the
XSPEC\footnote{http://heasarc.gsfc.nasa.gov/docs/xanadu/xspec/index.html}
X-ray spectral fitting package to perform this analysis. We fitted the
model to each difference spectrum by minimising the $\chi^{2}$
statistic. We calculated the modelled flux and 1-sigma errors on the
flux between 0.25 and 2.5\,keV for each EPIC-MOS instruments. The
keyword giving the area collected in the spectral extraction
(BACKSCAL) was converted into units of steradians and used to convert
the individual flux values to units of \fluns. Fluxes presented here
onwards are for a combined error weighted-average EPIC-MOS flux. The
error on the flux was calculated from the individual flux errors,
combined in quadrature. A histogram of the total spectrally fitted
flux for each of the SWCX set can be seen in
Figure~\ref{figfluxhist}. The minimum flux we observed for a SWCX case
was 2.2\,\fluns\ (observation id. 0112490301) and the maximum
50.1\,\fluns\ (observation id. 0085150301, \citet{carter2010}).

The relative strengths of the component lines to the SWCX spectral
model varied considerably within the SWCX set. Individual line fluxes
were calculated by finding the best fit model using all lines, as
described above, then setting other line normalisations to zero in
XSPEC. The flux for the individual ion species contribution was
calculated in the range 0.25 to 2.5\,keV.

\begin{figure}
  \includegraphics[width=0.475\textwidth, angle=0, bb=65 370 540 700]{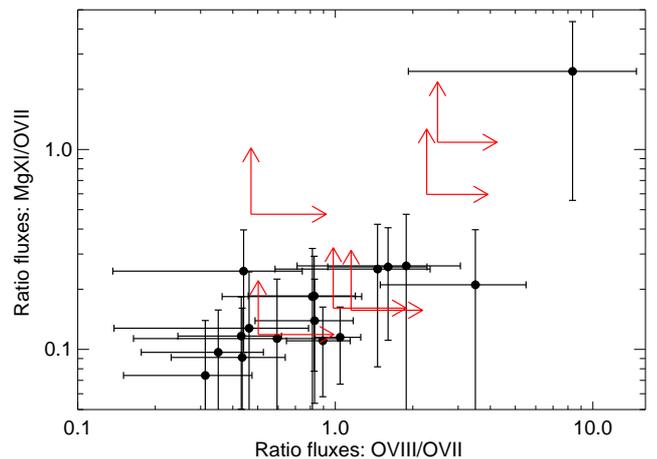}
  \caption{Ratio of \mage/\oxys\ to \oxye/\oxys\ where available for
    the SWCX set. Where appropriate we mark the lower limit (red).}
  \label{figmgxioxyratio}
\end{figure}

High \opluss\ to \oplussix\ and magnesium to oxygen ratios are used as
indicators of the presence of CME plasma \citep{zhao2009,
  richardson2004, zurbuchen}. \oxys\ is the dominant SWCX ion-species
in the majority of cases. In Figure~\ref{figmgxioxyratio} we plot the
ratio of the fluxes of the lines \mage/\oxys\ to \oxye/\oxys\ (using
those oxygen transitions available to us within our X-ray spectral
band). To look for plasma signatures with the highest charge states we
only plot those cases where the normalisation of the numerator in the
ratio is well constrained. Three observations are constrained to have
both a ratio of \mage/\oxys\ $>$ 0.6 and \oxye/\oxys\ $>$ 1.0. One of
these (with identifier 0085150301) was the observation previously
assigned to a passing CME and described in \citet{carter2010}. These
observations are therefore possible candidates for having observed CME
plasma with \xmm\ and are listed in Table~\ref{tabhighionratio}. We
quote the lower limit to the ratio in the case where the \oxys\ flux
is badly constrained, i.e. very weak. In this table we also quote the
mean value of the \opluss\ to \oplussix\ ratio during the period of
the observation, using values taken from the \ace\ SWICS
instrument. This data was only available in two of the cases in the
table. As \ace\ is found at Lagrangian point L1, we have time shifted
the solar wind data to account for the travel time to Earth, based on
the mean speed of the solar protons during the \xmm\
observation. Expected \opluss/\oplussix\ ratios for the slow and fast
solar wind are 0.27 and 0.03 respectively \citep{schwadron2000}. Both
of the observations with \ace\ SWICS data surpass both the nominal
slow and fast values by a considerable margin and would suggest that
\ace\ detected a CME plasma.  Although the identification of CME
plasma generally involves many more criteria to be satisfied, \xmm\
could provide supplementary spectral evidence to studies employing
in-situ dedicated solar wind monitors in the field of solar system
space science.

\begin{figure}
  \includegraphics[width=0.475\textwidth, angle=0, bb=80 370 540 700]{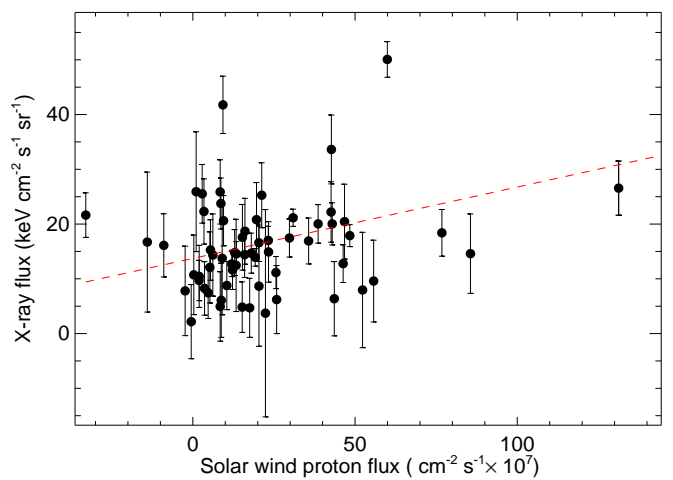}
  \caption{Observed flux versus mean solar wind proton flux. The red
    dotted line indicates a linear fit to the data.}
  \label{figoxysvsswflux}
\end{figure}

To test if any relationship exists between the flux of the SWCX lines
and increased solar wind flux, we plot in Figure~\ref{figoxysvsswflux}
the observed flux versus the difference in the mean solar wind proton
flux (as measured by \ace) between the SWCX-affected and SWCX-free
periods. We have again time-shifted the \ace\ data to account for the
distance between L1 and the Earth. Although there is considerable
scatter amongst these values, there is a positive correlation between
line flux and solar wind proton flux. We include in the plot the
linear fit as found by the IDL procedure \textit{linfit}. Proton flux,
for our SWCX set, is a good indicator of the presence of
SWCX-enhancement, if not the level of this enhancement. This is in
contrast with the results of \citet{henley2010}, whose SWCX cases were
considered to be due to SWCX occurring within the heliosphere and
therefore no correlation would be expected between an upstream solar
wind monitor and any SWCX enhancement. Heliospheric SWCX is expected
to vary on longer timescales than exospheric SWCX and therefore will
be harder to identify by the technique in this paper. However, at
certain times of the year \xmm\ may have a line-of-sight that passes
through the helium focusing cone, that could potentially produce a
variable signal in the line-band that may be detectable by this
technique (variations over a few hours, \citep{koutroumpa2007}). We
discuss this possibility further in Section~\ref{secmodvsxmm}. The
flux variations seen within our SWCX set are therefore due to local
X-ray emission in the vicinity of the Earth.

\begin{table}
  \caption{SWCX set observations exhibiting the highest
    \mage/\oxys\ and \oxye/\oxys\ ratios, or the lower limit (95\,\% confidence) to this ratio when \oxys\ is badly
    constrained. We also note the \ace\ SWICS mean value of the
    \opluss\ to \oplussix\ ratio when available, with the standard
    deviation of this ratio given as the error.}
\begin{tabular}{lllll}
  \hline\hline
  Revn & Obsn       & \mage/ & \oxye/ & Mean \\ 
       &            & \oxys  & \oxys  & \opluss/\oplussix \\ 
  \hline
  0342 & 0085150301 & 2.5$\pm$1.9  & 8.3$\pm$6.4  & 0.58$\pm$0.54 \\
  0494 & 0109120101 & $\ge$0.60    & $\ge$2.26    & 1.53$\pm$0.83 \\
  0747 & 0200730401 & $\ge$1.09    & $\ge$2.49    & \ldots \\
\end{tabular}
\label{tabhighionratio}
\end{table}

\section{Example cases of SWCX enhancement}\label{seccases}

\begin{figure*}
  \centering
  \begin{tabular}{c}
    \includegraphics[width=.45\textwidth, angle=0, bb=80 380 525 705]{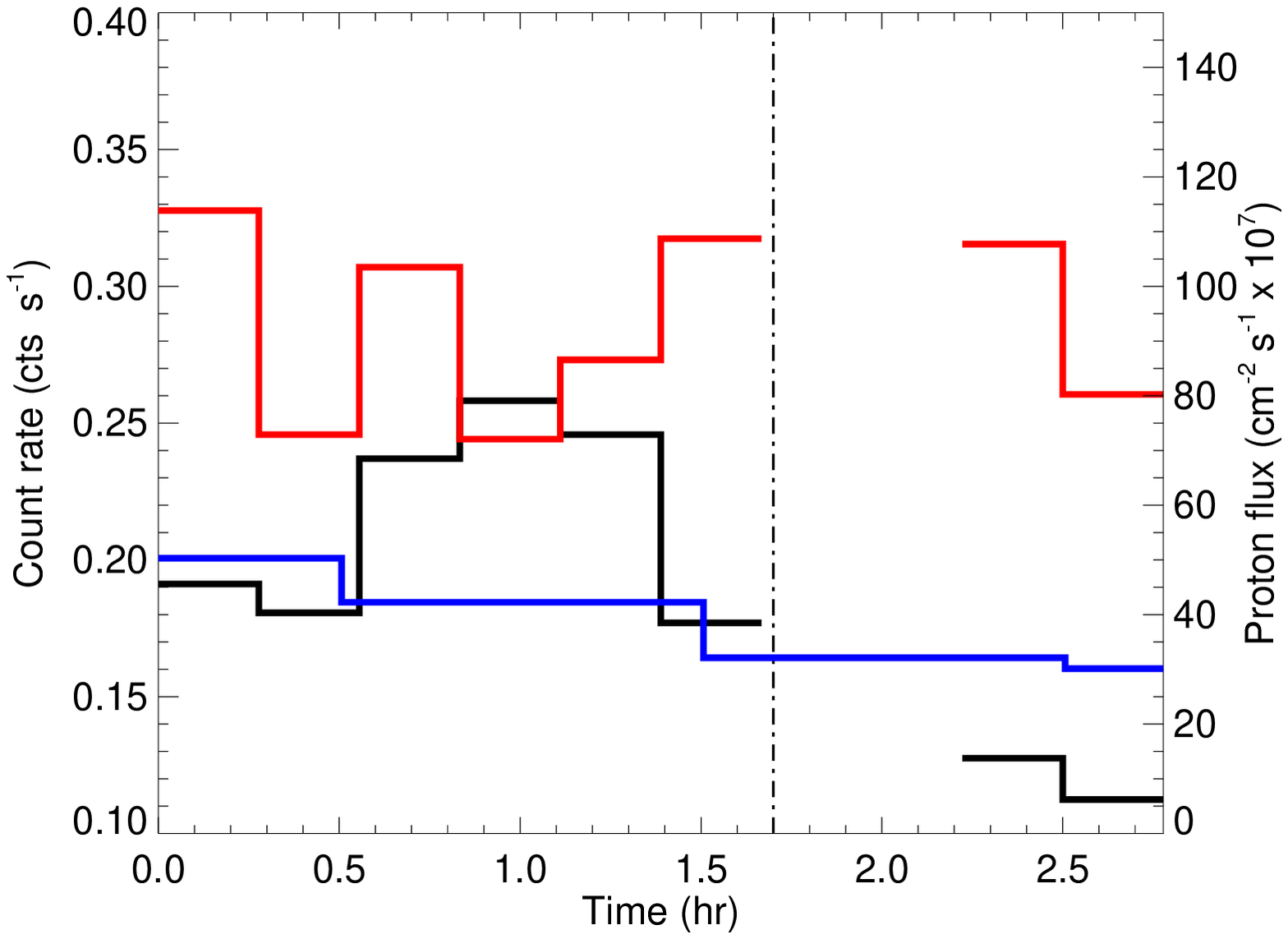} 
    \includegraphics[width=.45\textwidth, bb=0 0 683 500]{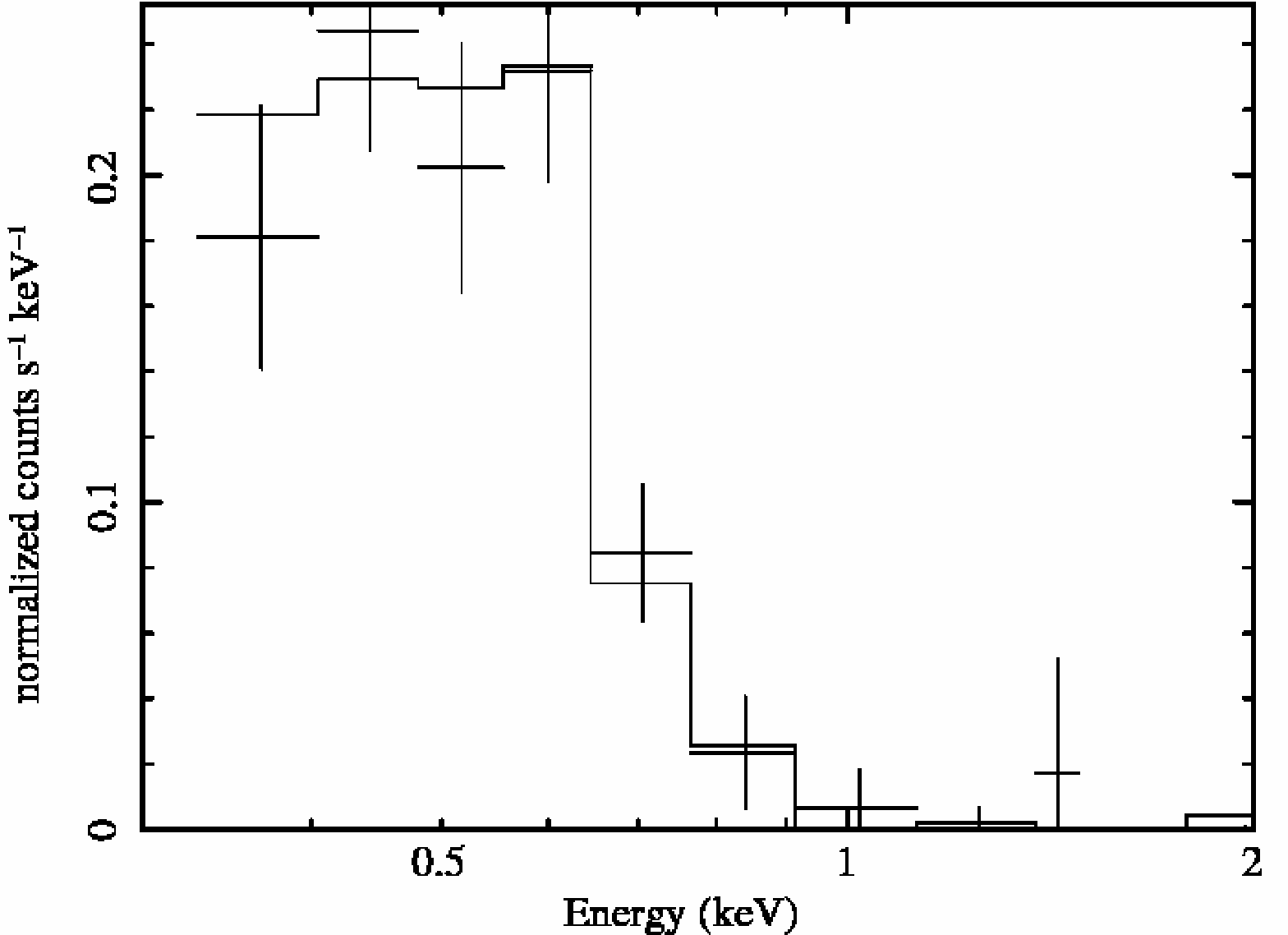}\\
    \includegraphics[width=.45\textwidth, angle=0, bb=80 380 525 705]{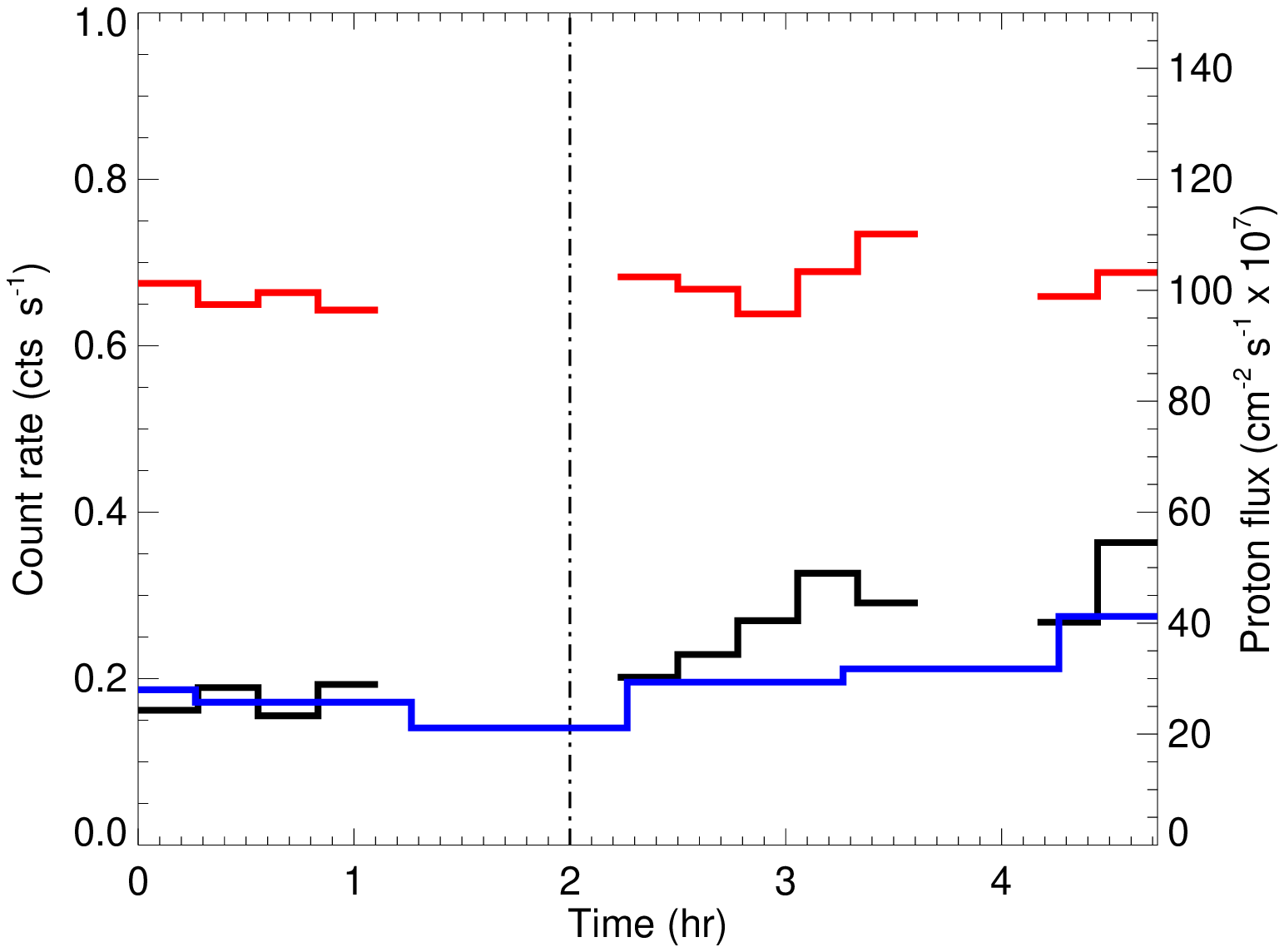}
    \includegraphics[width=.45\textwidth, bb=0 0 683 500]{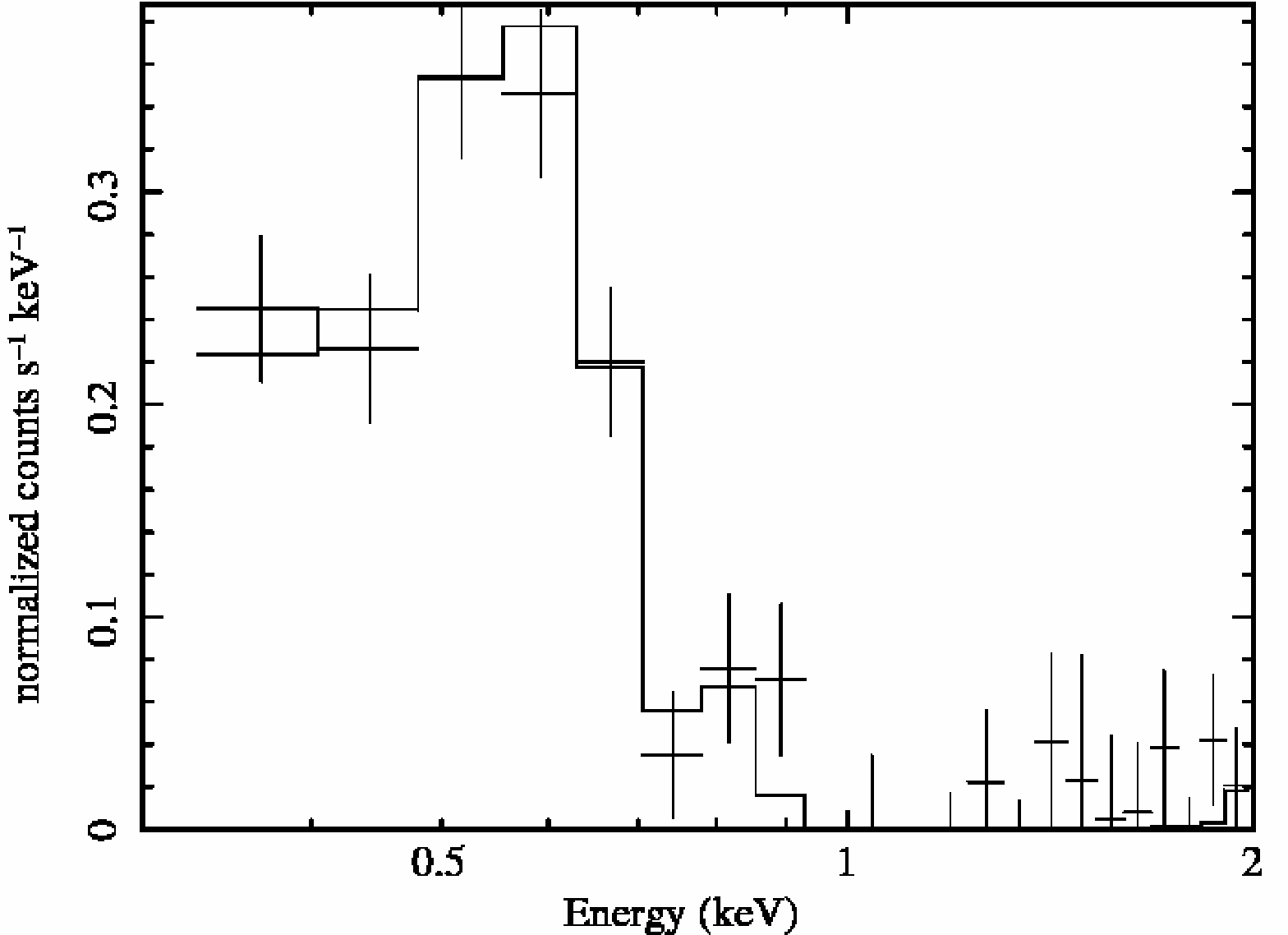}\\
    \includegraphics[width=.45\textwidth, angle=0, bb=80 380 525 705]{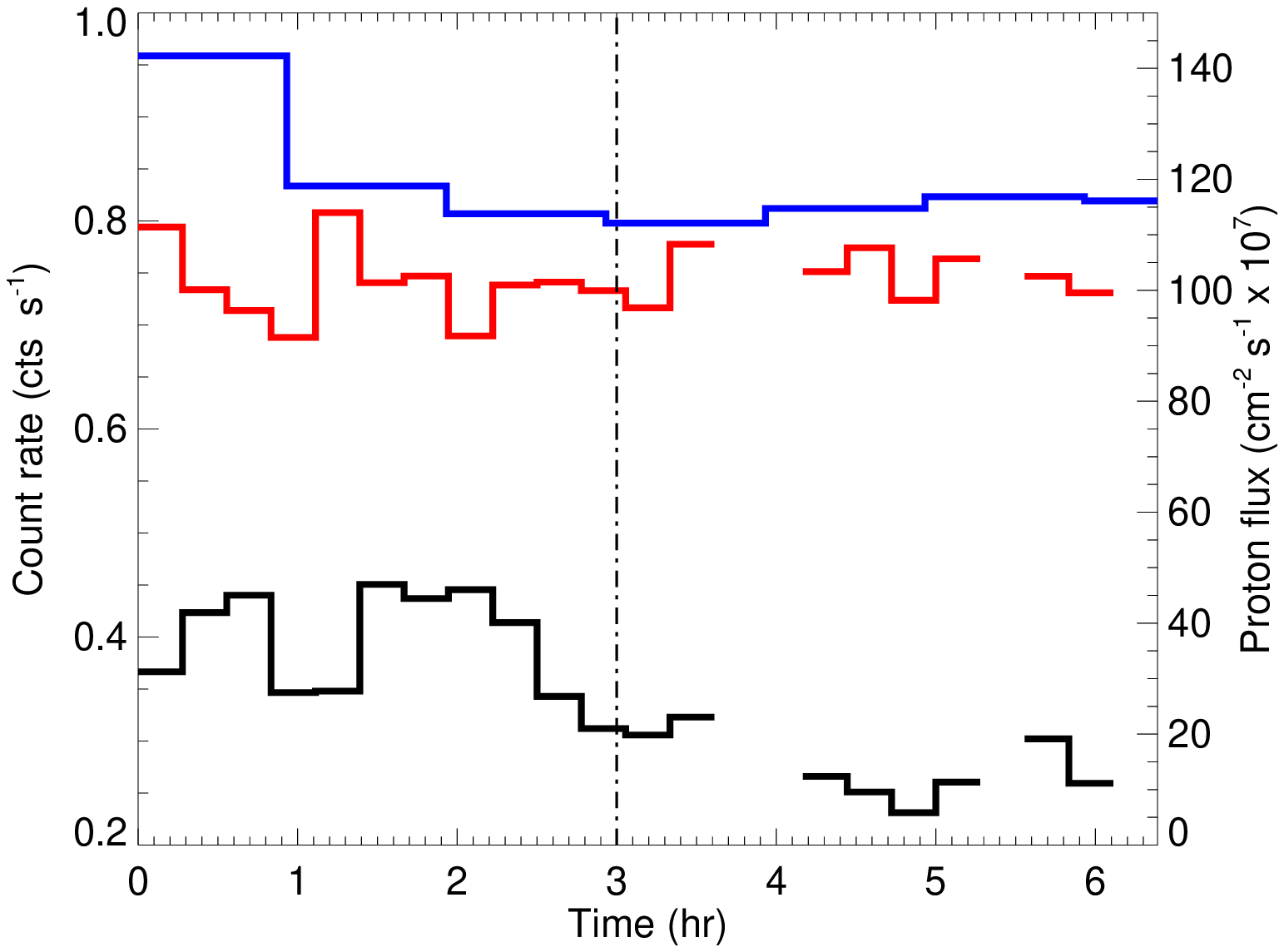}
    \includegraphics[width=.45\textwidth, bb=0 0 683 500]{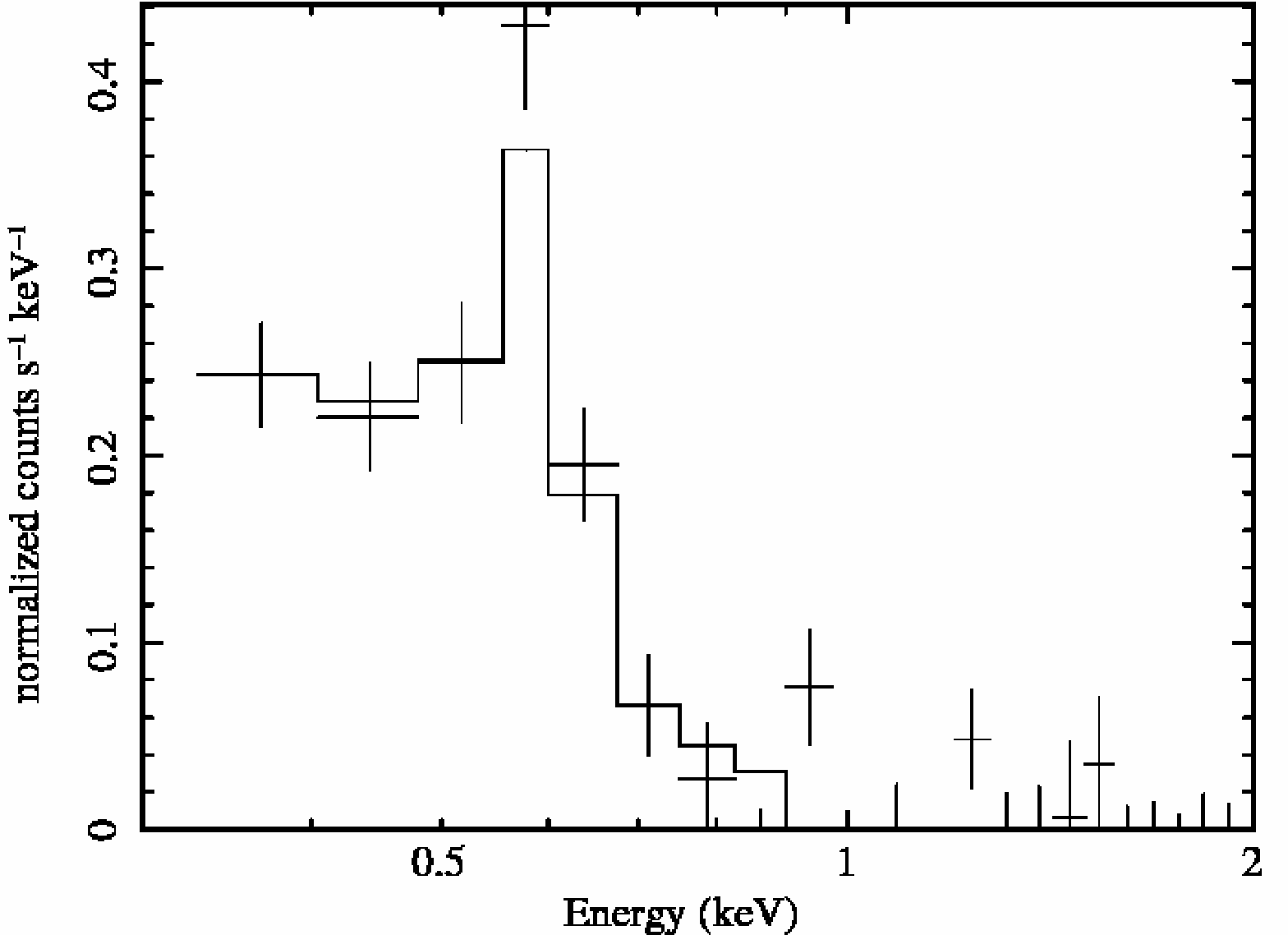}
\end{tabular}
\caption{Example cases of SWCX enhancement, for (top row) observation
  id. 0150610101, (middle row) observation id. 0054540501 and (bottom
  row) observation id. 0113050401. Each left-hand panel shows the
  line-band (black) and continuum band (red) non-mean-adjusted
  lightcurves along with the solar proton flux (blue, right-hand
  y-axis). The split between the SWCX-free and SWCX-affected period is
  indicated by the dashed vertical line. In the right-hand panels we
  show the combined EPIC-MOS difference spectrum and the model fitted
  to the data for each case (solid line).}
  \label{figswcxexam}
\end{figure*}

In this section we comment on the three newly identified exospheric
cases from Table~\ref{tabtop}. Lightcurves for each, over-plotted with
the solar proton flux as recorded by \ace, are given in
Figure~\ref{figswcxexam}. The solar proton lightcurves have been
adjusted for the distance between \ace\ and the Earth by adding a
delay based on the distance to \ace\ and the mean solar proton speed,
assuming a planar wavefront travelling on the Sun-Earth axis. We also
plot a difference spectrum for each observation by combining data from
both EPIC-MOS cameras and over-plot the best fit model to the data, as
described in Section~\ref{secspec}. Individual fluxes for a selection
of prominent lines are given for each case in
Table~\ref{tabexampleions}.

\begin{itemize}

\item Observation 0150610101 (revolution 0623)

  The line-band lightcurve shows a period of enhanced count rate at
  the beginning of the observation. The \ace\ solar proton flux is
  raised at the beginning of the lightcurve and reduces as the
  lightcurve progresses. The difference spectrum exhibits emission at
  \oxys, \oxye, along with evidence of carbon emission below
  0.5\,keV. The flux observed between 0.25 and 2.5\,keV was
  20.6\,\fluns.

\item Observation 0054540501 (revolution 0339)

  The line-band lightcurve shows an enhancement during the latter part
  of the observation, which is also observed in the \ace\ solar proton
  flux. The difference spectrum exhibits emission in the oxygen band,
  along with evidence of carbon emission below 0.5\,keV. The flux
  observed between 0.25 and 2.5\,keV was 7.9\,\fluns.

\item Observation 0113050401 (revolution 0422)

  The line-band lightcurve shows a period of enhanced count rate at
  the beginning of the observation. A short enhancement period is seen
  in the \ace\ solar proton flux and the overall magnitude of this flux
  is much higher than the other two cases in this section. The flux
  observed between 0.25 and 2.5\,keV was 25.9\,\fluns.

\end{itemize}

\begin{table*}
  \caption[]{Most prominent ion line fluxes for example cases. Fluxes are quoted in units of \fluns. Upper limits (95\,\% confidence) are given for very weak lines.}
    \begin{tabular}{llllll}
\hline\hline
Revn & Obsn & \cs & \nsix & \oxys & \oxye \\
\hline 
0623 & 0150610101 & 4.86$\pm$1.10 & 2.19$\pm$1.48 & 7.73$\pm$0.37 & 2.42$\pm$0.61 \\
0339 & 0054540501 & 5.61$\pm$0.85 & $\leq$3.50 & 11.73$\pm$0.27 & 4.74$\pm$0.41 \\
0422 & 0113050401 & 9.05$\pm$0.41 & $\leq$2.80 & 11.60$\pm$0.27 & 4.36$\pm$0.46 \\
      \end{tabular}
      \label{tabexampleions}
\end{table*}

\section{Modelling the expected emissivity}\label{secemiss}

The expected X-ray emissivity of SWCX emission from the solar wind
interaction with the magnetosheath can be estimated from the
integrated emission along the line-of-sight for the observer. We have
developed a model, applicable to local interplanetary space, to
calculate this emission. We have not attempted to include any
contribution from further into the heliosphere, as to increase the
integration length would lead to greater uncertainty in the underlying
parameters of the solar wind on large spatial scales. We assume that
the solar wind parameters used in the model are approximately constant
(excluding the magnetosheath region) along the line-of-sight. The
emissivity expected \citep{cravens2000} is given by the expression:

\begin{equation}
P_{^\chi} = \alpha\eta_{^{SW}}\eta_{^n}\langle g \rangle\ eV cm{^{-3}}s{^{-1}}
\label{eqnemiss}
\end{equation}

where $\alpha$ is the efficiency factor dependent on various aspects
of the charge exchange such as the interaction cross-section and the
abundances of the solar wind ions, $\eta_{^{SW}}$ is the density of
the solar wind protons, $\eta_{^n}$ is the density of the neutral
species and $\langle g \rangle$ is their relative velocity.

For each observation under study in paper II, we wish to test whether
any relationship exists between the total SWCX flux seen and the
theoretical integrated X-ray emission along the line-of-sight (based
on Equation \ref{eqnflux}, \citet{cravens2000}).

\begin{equation}
F = \frac{1}{4\pi} \int_0^\infty P_{^\chi} ds\ eV\ cm{^{-2}}s{^{-1}}sr{^{-1}} 
\label{eqnflux}
\end{equation}

We take data describing the conditions in the solar wind from \ace\
(\textit{Level2} processed data, merged instrument data using hourly
averages) at the time of each observation. We needed to apply a delay
to the signal received, to account for the separation between \ace\
and the Earth. This delay will be time variable and will depend on the
speed and orientation of the solar wind. However, as a first
approximation, we have taken the average solar proton speed of the
data and assumed a planar wavefront travelling anti-sunward
perpendicular to the GSE$\mhyphen$X axis. We calculate the delay
required for the wavefront to travel from \ace\ to the Earth.

Throughout this work we assume a geocentric solar-ecliptic coordinate
system (GSE), where positive X is directed from the Earth to the Sun, positive Y
opposes planetary motion and positive Z is parallel to the direction
towards the north ecliptic pole.

Then for each time bin of an observation:

\begin{itemize}
\item We extract the solar wind proton velocity and temperature from
  the \ace\ data, and for these parameters we calculate a thermal
  velocity and average speed, using Equations \ref{eqnthermv} and
  \ref{eqnavev}.

\begin{equation}
\nu_{th} = \sqrt{3k_{b}T}
\label{eqnthermv}
\end{equation}

\begin{equation}
\langle g \rangle = \sqrt{\nu_{th}^{2}+u_{sw}^{2}}
\label{eqnavev}
\end{equation}

\item We estimate the position of the magnetopause, based on the model
  of \citet{shue1998}. To do this we use information regarding the
  strength and direction of the interplanetary magnetic field ($B_{z}$
  component). We currently assume that the magnetopause shape is
  symmetrical about the GSE$\mhyphen$X axis and place the magnetopause
  standoff distance along this axis.
\item Using the magnetopause location as a guide, we approximate the
  position of the bow shock. We base the shape on a simple parabola
  and calculate the bow shock standoff distance using the solar wind
  pressure and the relationship in \citet{khan1999}. The magnetosheath
  and bowshock together define the magnetosheath region.
\item We create an Earth-centric square image for use in subsequent
  steps. The side length of the image is 200\,\er. This image is
  divided into cells, with side length 0.5\,\er. The magnetopause
  shape is projected onto this image. We are able to use an image
  rather than a cube due to the assumption made previously regarding
  the symmetry of the magnetosheath shape about the GSE$\mhyphen$X
  axis.
\item We find the neutral density of hydrogen atoms for each cell. We
  use the \citet{ostgaard2003} model for neutral hydrogen density
  profiles around the Earth, but limit this to a minimum density of
  0.4\,$\mathrm{cm}^{-3}$ \citep{fahr1971}.
\item We determine the line-of-sight of the \xmm\ pointing through the
  grid by extracting the relevant information from the ODF and
  converting the positions and target pointing direction to the GSE
  coordinate system.
\item We find the velocity and density of the solar wind for each
  cell, (\citet{spreiter1966}, K.D. Kuntz private communication). As
  the solar wind passes the bowshock and enters the magnetopause its
  density increases (by about a factor of four in the subsolar region,
  as compared to the unperturbed value), and the velocity drops to
  about one tenth.
\item The value of $\alpha$ is dependent on the abundances of the ion
  species contributing to the charge exchange process, along with the
  cross-section and energy of each interaction with the neutral donor
  in the energy band of interest. The neutral donor is hydrogen in the
  geocoronal case. The relative abundances found in the solar wind
  vary considerably with solar wind state. The composition of the
  solar wind generally follows abundances seen in the photosphere, but
  can vary by up to a factor of about 2 (fast wind) or 4 (slow wind)
  for elements with first ionisation potential (FIP) below the
  Lyman-$\alpha$ limit of 10.2\,eV \citep[and references
  therein]{richardson2004}. However, we use the slow solar wind
  abundances for an ion species with respect to oxygen, as listed in
  \citet{schwadron2000}. We use an oxygen to hydrogen ratio of 1/1780
  for solar wind speeds of $\le 650\,\mathrm{km}\,\mathrm{s}^{-1}$ or 1/1550 for speeds
  above this threshold. For this modelling we consider contributions
  to the emission from the principal and minor transitions as
  described in Section~\ref{secspecmodel}. Cross-sections for charge
  exchange transitions are dependent on solar wind speed. We calculate
  an $\alpha$ map with the same dimensions and binning as that of the
  Earth grid and populate this map with values of $\alpha$ depending
  on the speed of the solar wind, unperturbed outside of the
  magnetosheath or perturbed inside the magnetosheath as described
  above.
\item We multiply the solar wind velocity, solar wind density and
  neutral hydrogen density together for each of the cells in the
  line-of-sight and multiply this value by the efficiency factor
  $\alpha$ for each cell. This is the emissivity of each cell.
\item We sum all cells in the line-of-sight, accounting for the number
  of cells included in the integral, to give the value of the
  emissivity metric, approximating Equation~\ref{eqnflux}.
\end{itemize}

There are some known limitations to this model, such as:

\begin{itemize}
\item There are no magnetosheath cusps (increased density or
  modifications to the velocity of the solar wind specific to these
  regions) included in the Spreiter approximation
\item The neutral hydrogen has been modelled as spherically
  symmetrical about the Earth. However, there may be density
  enhancements in regions of the exosphere.
\item The abundances of the solar wind are not constant, but will
  change for example with the phase of the solar cycle or the injection of
  plasma from a CME.
\item The magnetopause and bow shock stand-off distances have been
  assumed to be on the GSE$\mhyphen$X axis, this may not be the case.
\item The interstellar neutral density may be significantly different
  from that of the approximate limiting density applied in this model.
\end{itemize}

We also consider an adapted model, whereby the $\alpha$ calculated is
dependent on the relative abundances of \oplussix\ and \opluss. We use
the ratio of these ionisation states, taken from the \ace\ merged,
hourly-averaged data sets to re-calculate the abundance of \opluss\
assuming the initial \oplussix\ and \opluss\ abundances as found in
\citet{schwadron2000}. We then re-calculate the value $\alpha$ and the
subsequent line-of-sight flux. These results will be referred to as
the \modtwo\ results and will be discussed in
Section~\ref{secmodvsxmm}.

\subsection{Modelled emission results}\label{secmodvsxmm}
\begin{figure}
  \includegraphics[width=.475\textwidth, angle=0, bb=68 362 536 656]{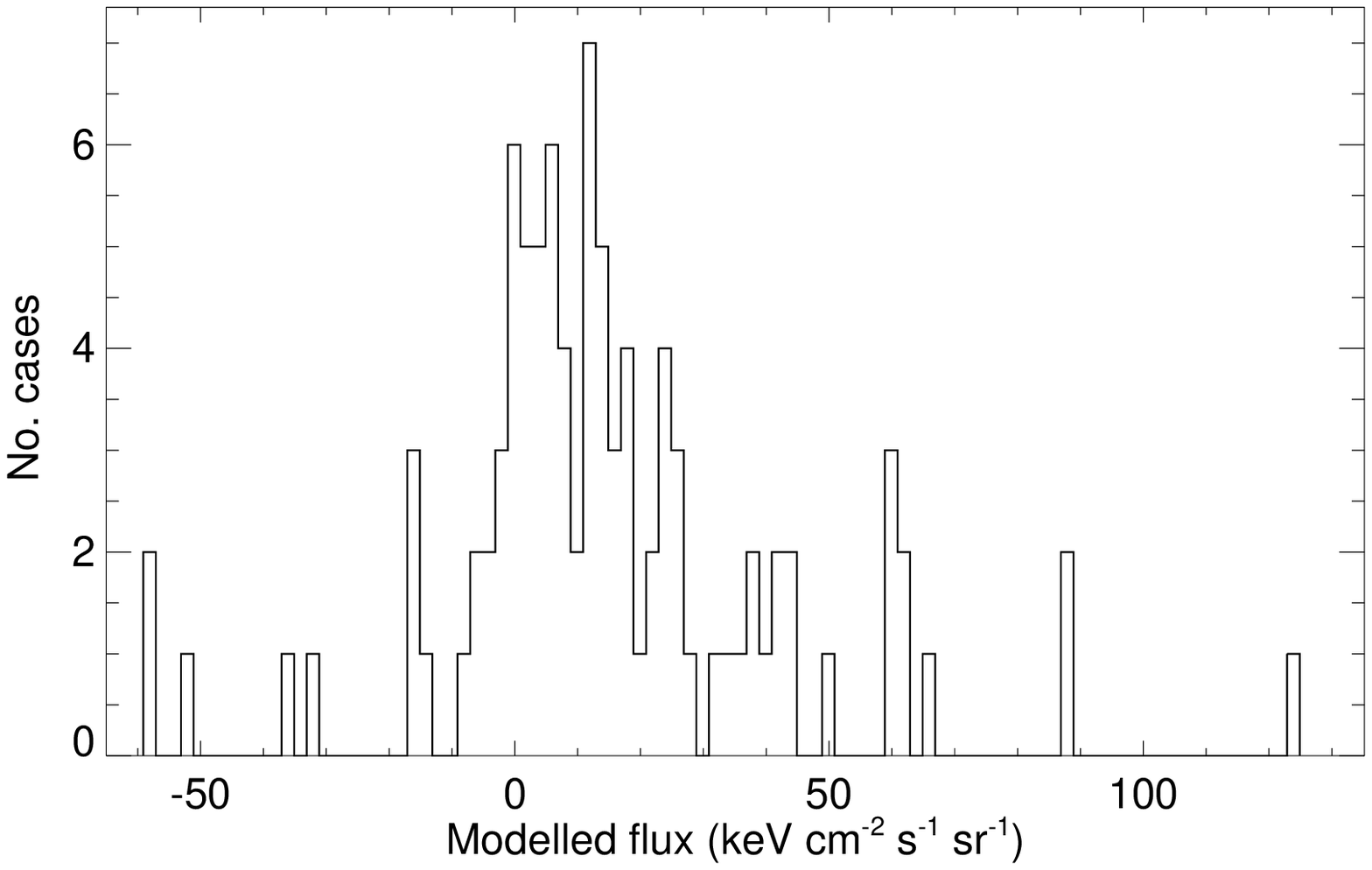}
  \includegraphics[width=.475\textwidth, angle=0, bb=84 370 538 698]{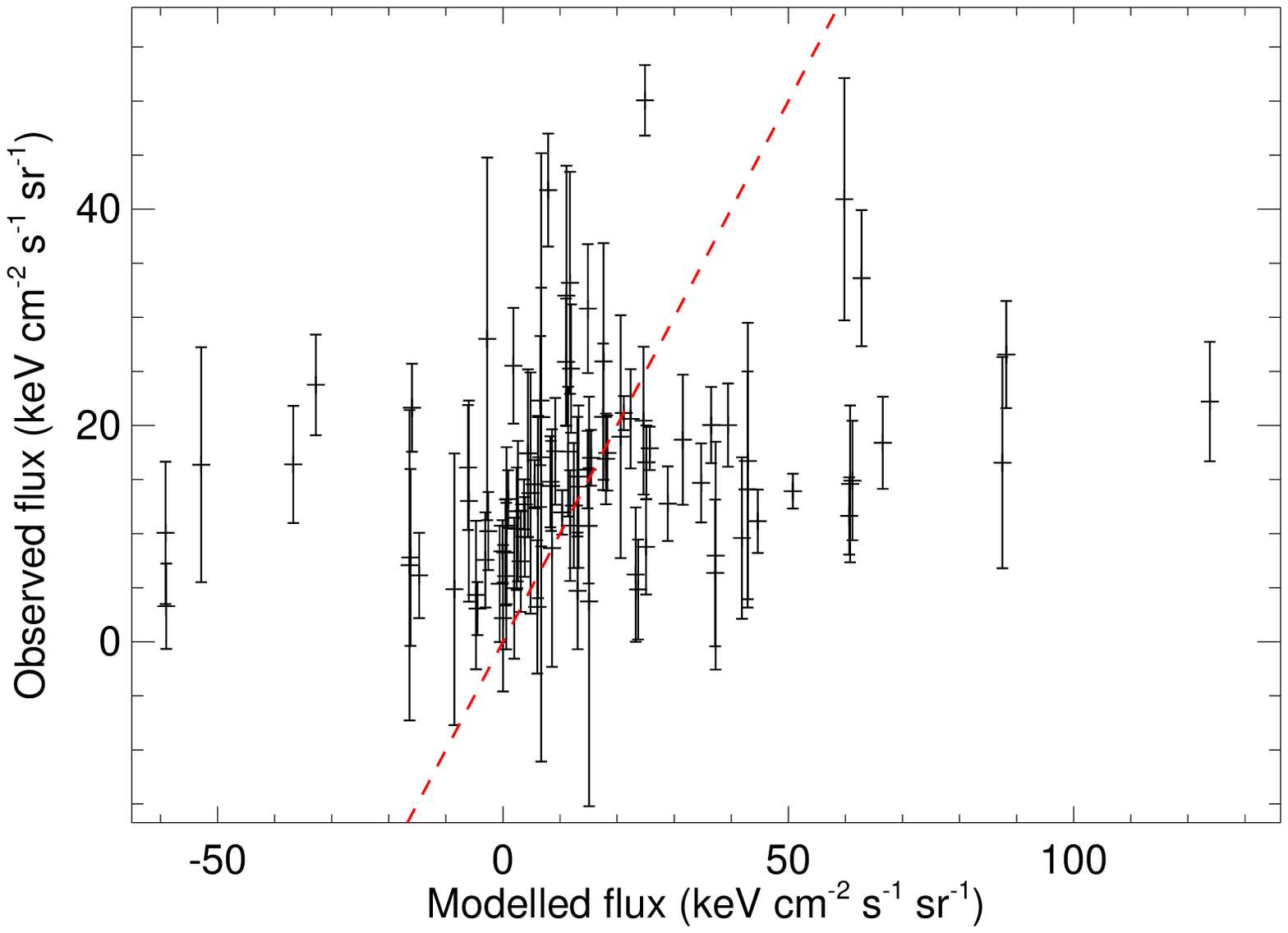}
  \caption{Top: histogram of the modelled fluxes. Bottom: observed
    flux (0.25 to 2.5\,keV) versus the modelled flux for the SWCX
    set. A line (dashed, red) of gradient unity has been added to the
    graph to aid the eye.}
  \label{figobsvsmeafl}
\end{figure}

Modelled lightcurves were produced for each SWCX case when there were
data available from \ace. We split the modelled emission based on the
time periods used for the creation of the spectra in
Section~\ref{secspecmodel}. The resultant flux is the difference
between the mean modelled flux during the SWCX-affected and the
SWCX-free periods. A histogram of the modelled fluxes is given in
Figure~\ref{figobsvsmeafl} (top panel), along with a scatter plot
showing the observed flux versus the modelled resultant flux for each
exospheric-SWCX case (bottom panel). In general there is a positive
correlation between the modelled and observed flux. For a few cases
the modelled flux is negative. This happens when the SWCX-affected
period, as determined using an enhancement seen in the observed
line-band lightcurve, occurred in the opposite period to the maximum
expected modelled flux. The enhancement in the observed line-band
lightcurve occurred sufficiently far away in time from any peak seen
in the modelled X-ray flux lightcurve.

\begin{figure*}
  \centering
  \includegraphics[width=.33\textwidth, angle=0, bb=60 360 530 700]{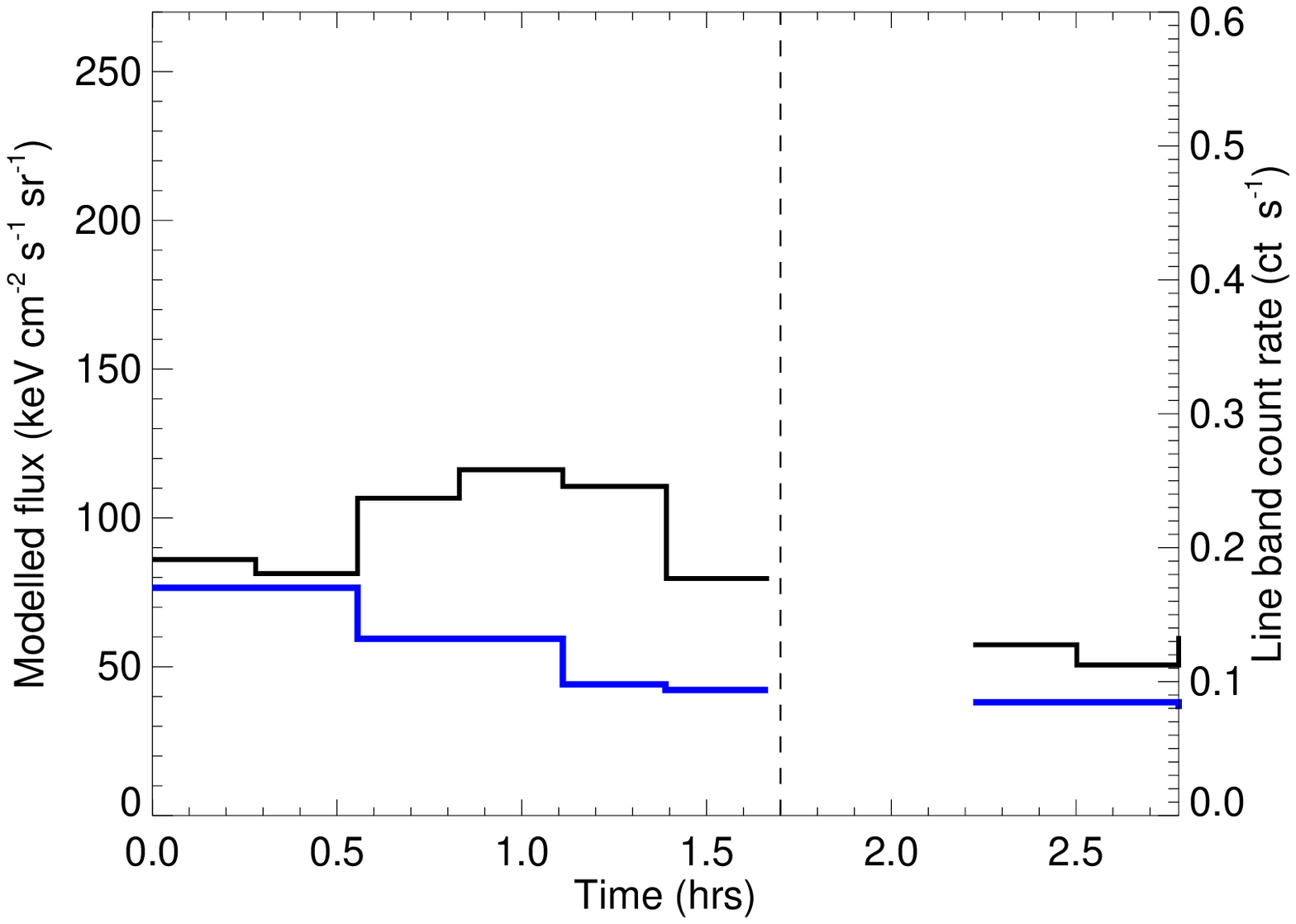}
  \includegraphics[width=.33\textwidth, angle=0, bb=60 360 530 700]{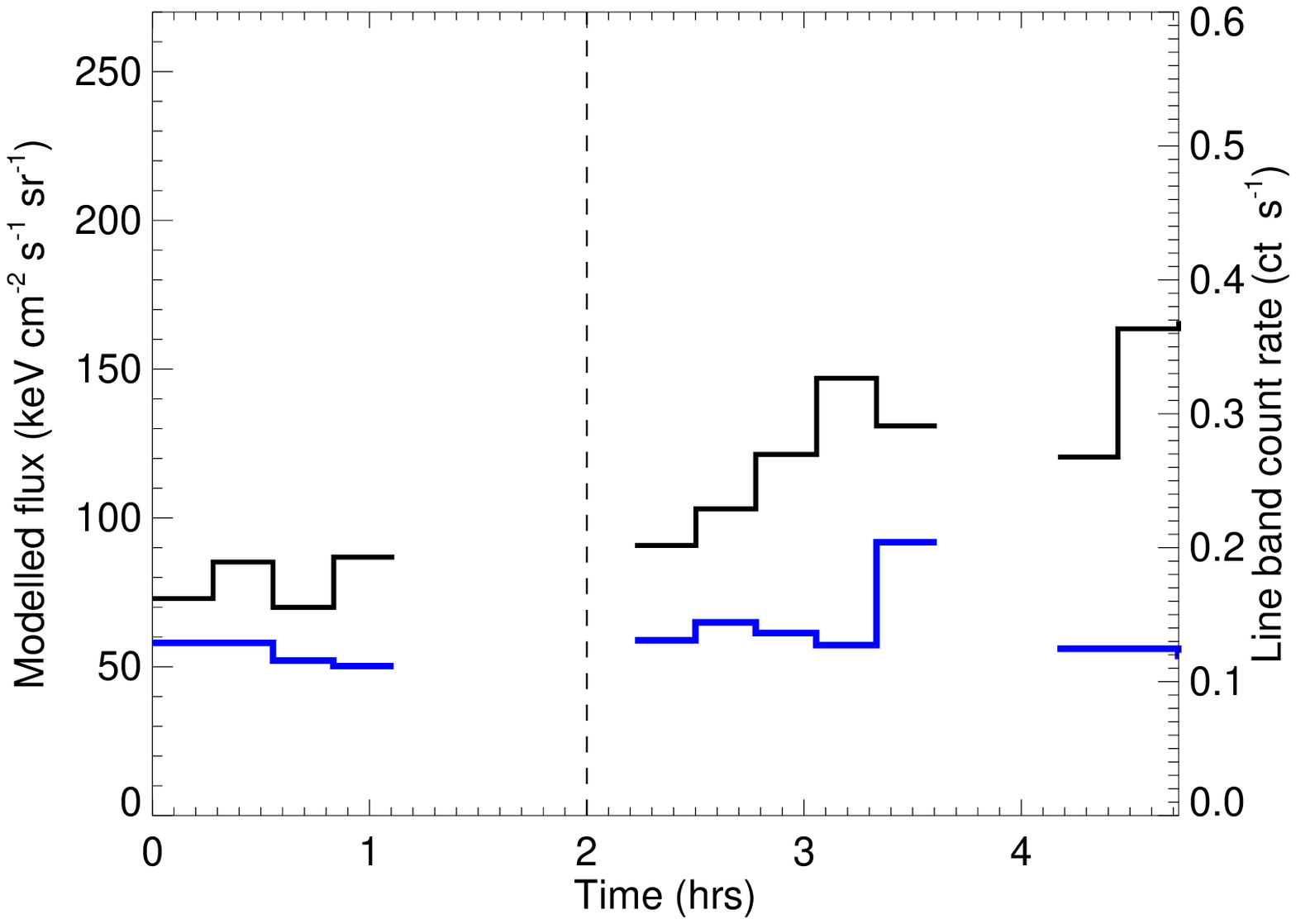}
  \includegraphics[width=.33\textwidth, angle=0, bb=60 360 530 700]{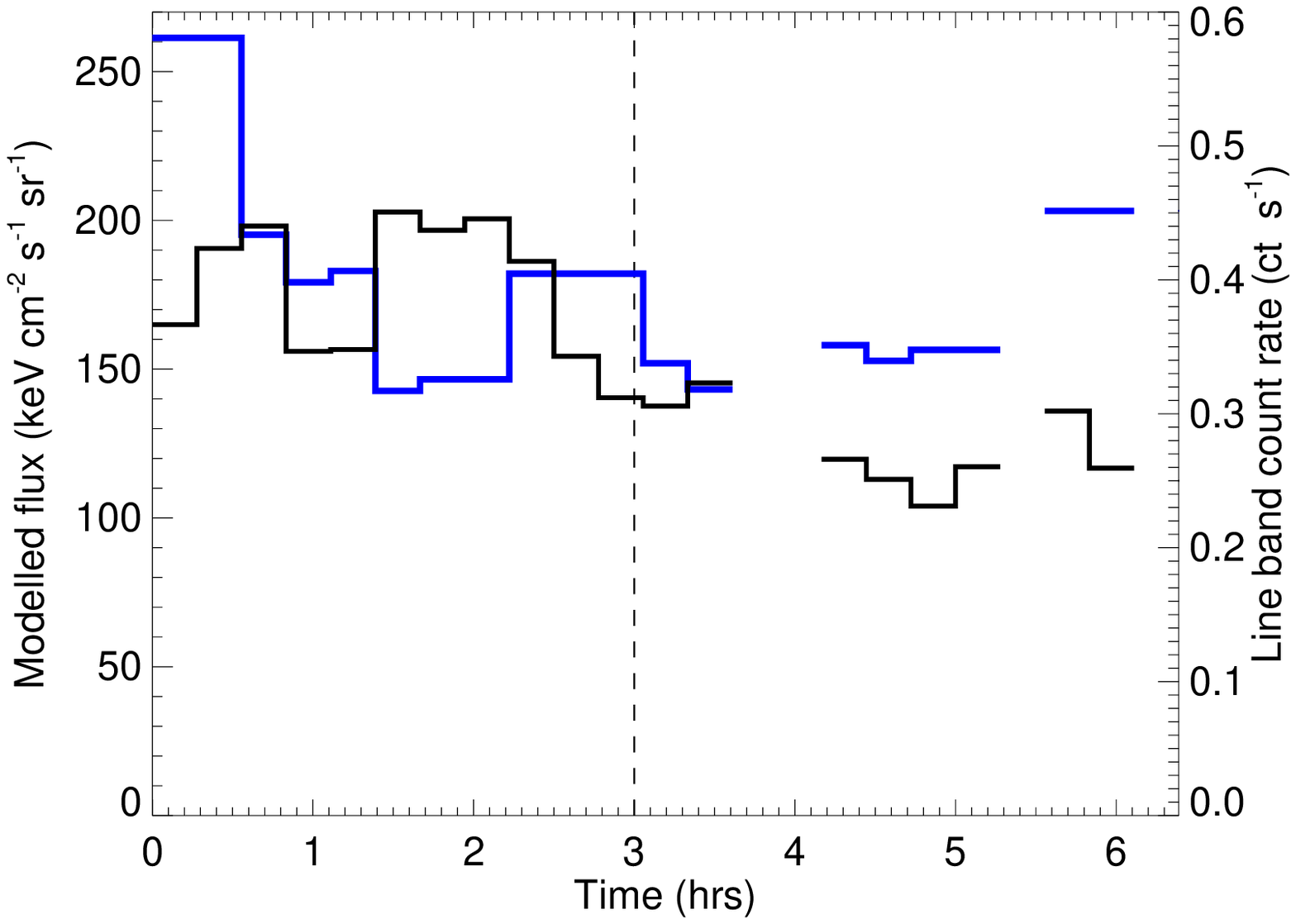}
  \includegraphics[width=.33\textwidth, angle=0, bb=60 360 530 700]{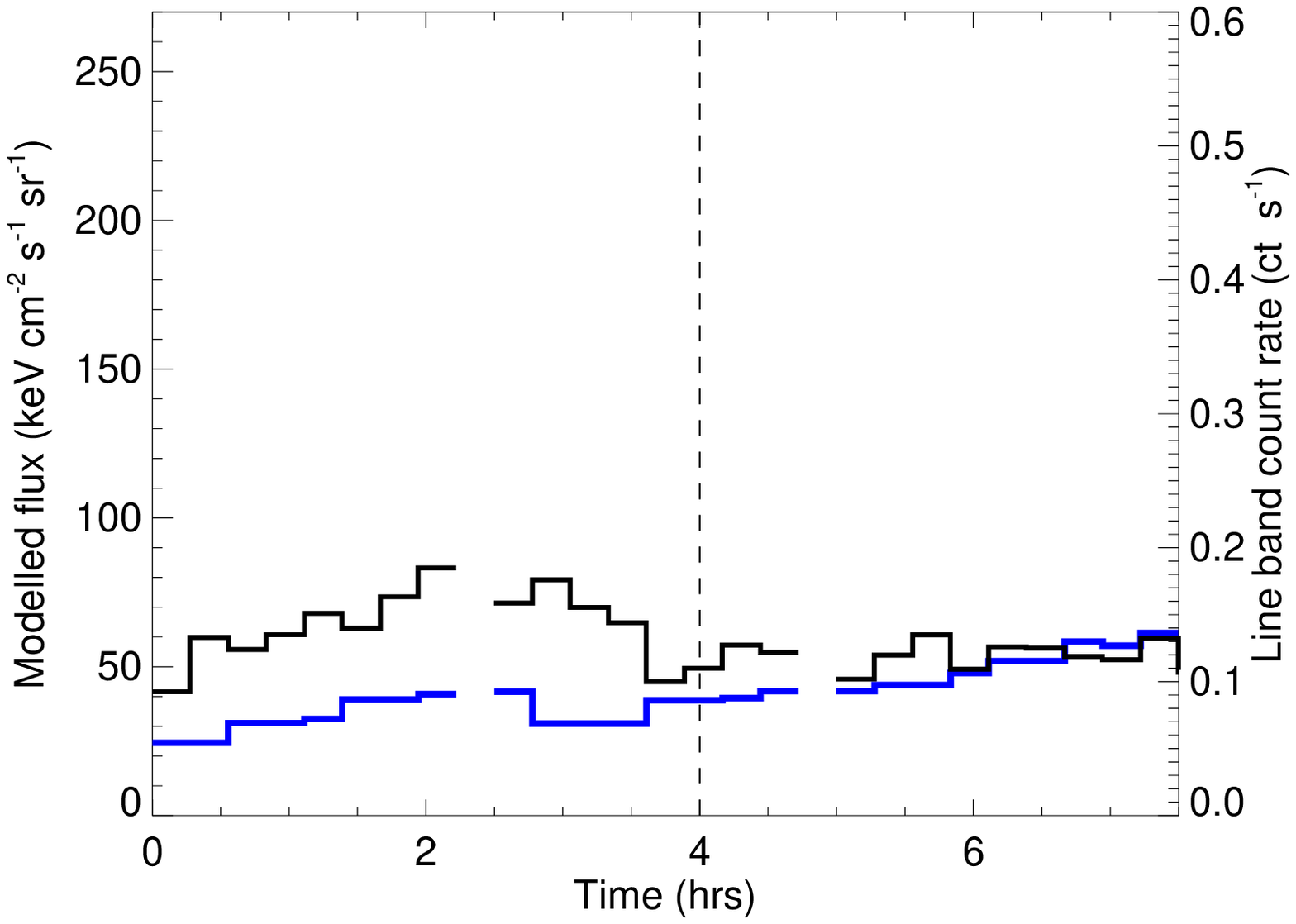}
  \includegraphics[width=.33\textwidth, angle=0, bb=60 360 530 700]{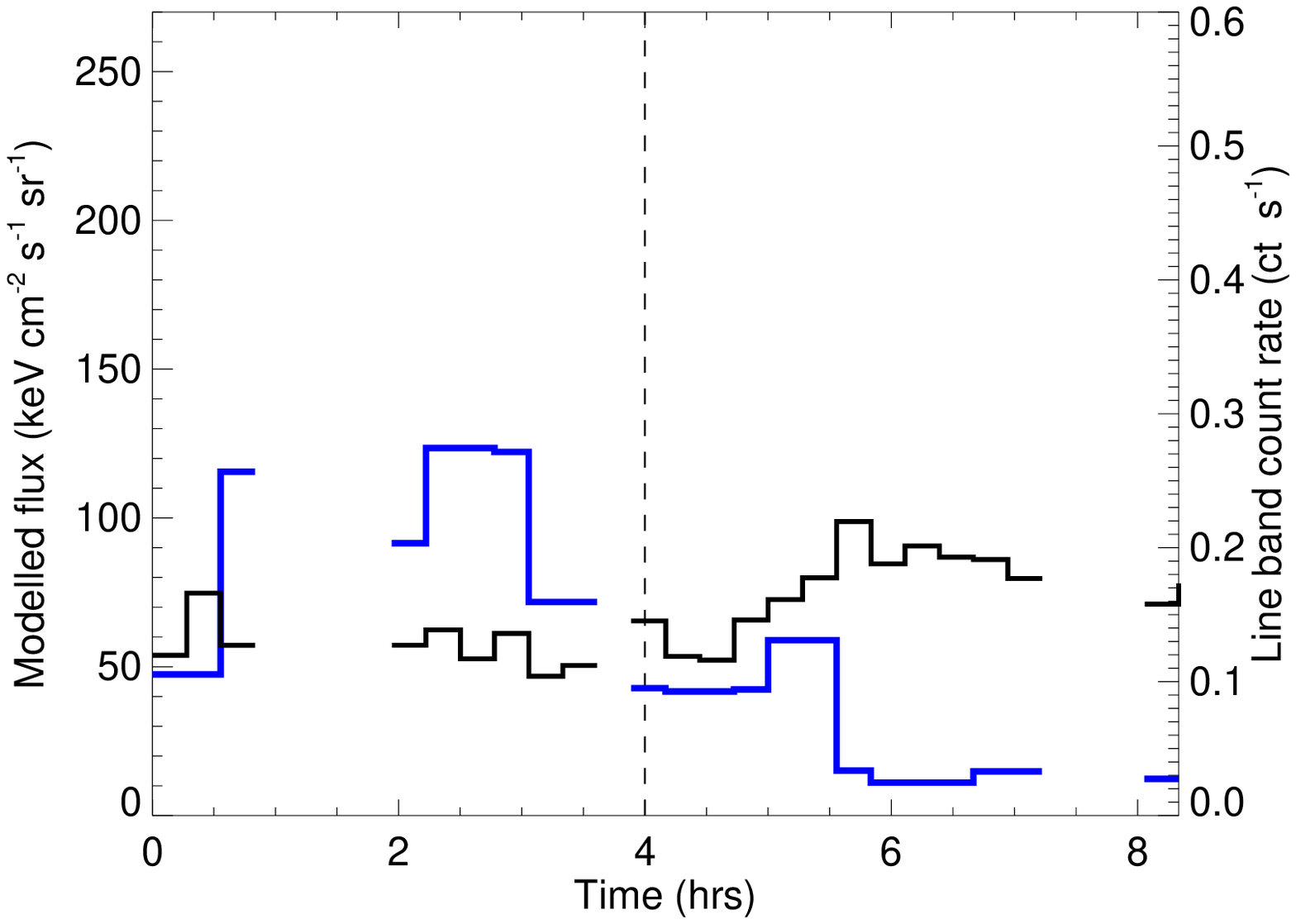}
  \includegraphics[width=.33\textwidth, angle=0, bb=60 360 530 700]{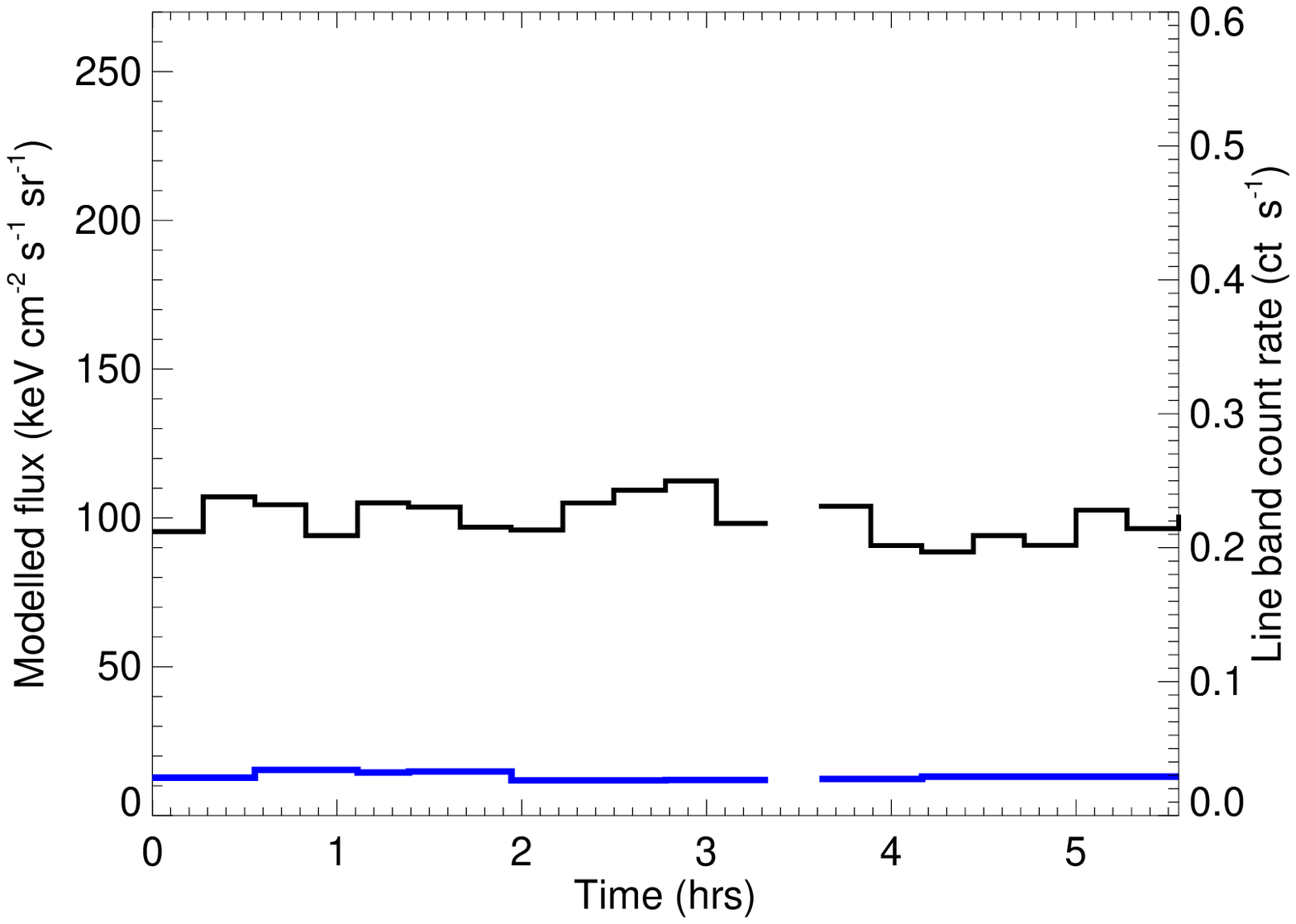}
  \caption{Example modelled (blue, in \fluns, left-hand y-axis) and
    \xmm\ line-band (black, $\mathrm{ct}\,\mathrm{s}^{-1}$, right-hand y-axis)
    lightcurves. Observations with identifiers (top-left) 0150610101,
    (top-middle) 0054540501, (top-right) 0113050401 show the model
    lightcurve generally following the shape of the \xmm\
    lightcurve. Observations (bottom-left) 0141150101 and
    (bottom-middle) 0150320201 show the modelled lightcurve peak in a
    different period to the \xmm\ lightcurve and (bottom-right)
    0301410601 is an example from an observation without a SWCX
    enhancement. Five panels show the split between the SWCX-affected
    and SWCX-free periods (vertical dashed line).}
  \label{figexamplemod}
\end{figure*}

In Figure~\ref{figexamplemod} (top-row) we show modelled lightcurves
for the three top new cases of Table~\ref{tabtop}. Contributions from
the model were only taken for the periods where there were counts in
the \xmm\ lightcurves (periods not removed during the filtering
process). Example lightcurves of cases where the modelled flux is
negative are given in the first two panels of the second row of
Figure~\ref{figexamplemod}. We also plot a modelled lightcurve when
SWCX was not detected (below the thresholds for \redc\ and \cratio) in
the bottom-right panel of the same figure. In this case the line-band
lightcurve does not vary significantly. The modelled emission in this
case is small compared to the SWCX cases presented in the other
examples.

\begin{figure*}
  \includegraphics[width=.475\textwidth, angle=0, bb=50 370 538 698]{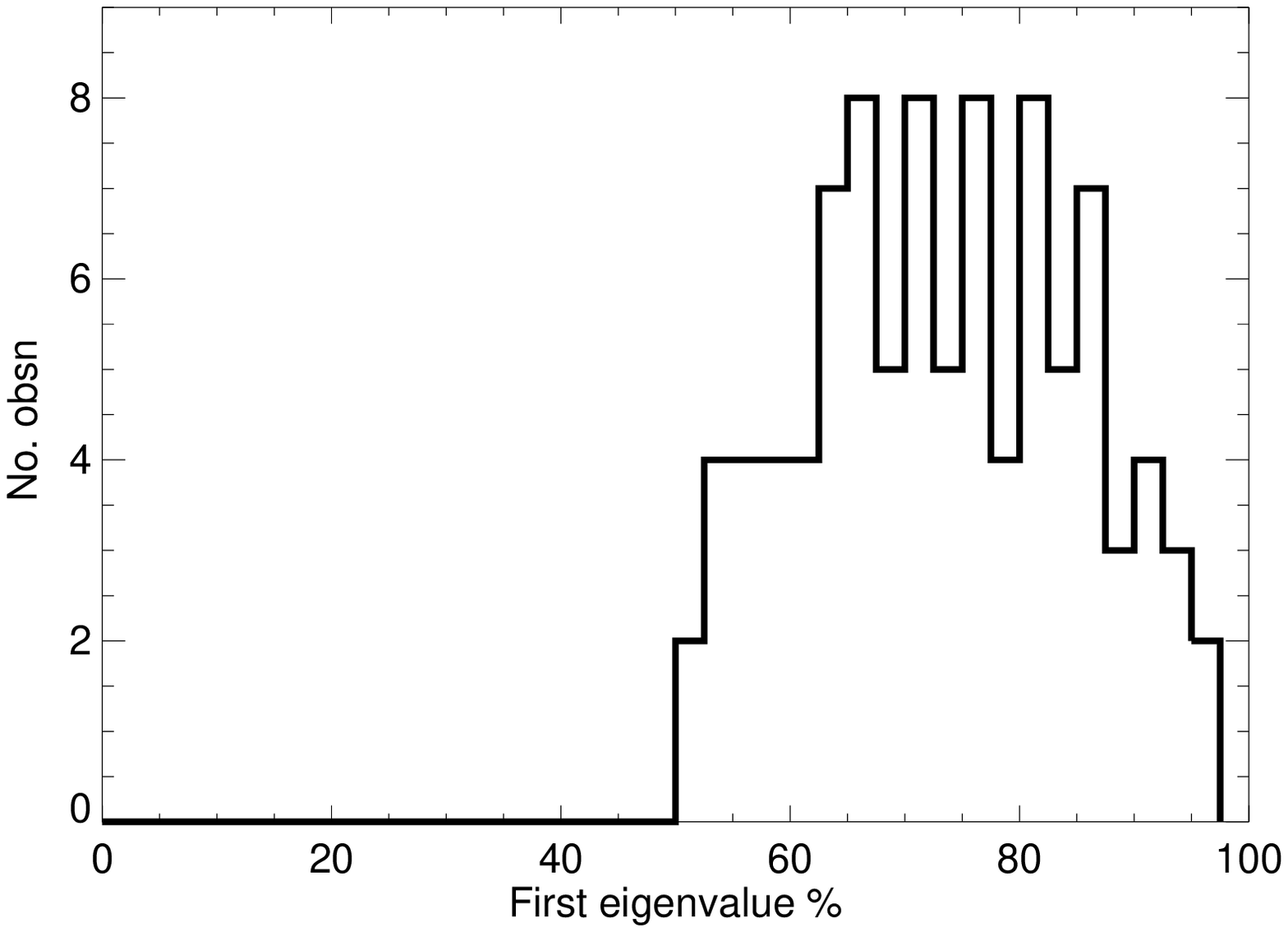}
  \includegraphics[width=.475\textwidth, angle=0, bb=50 370 538 698]{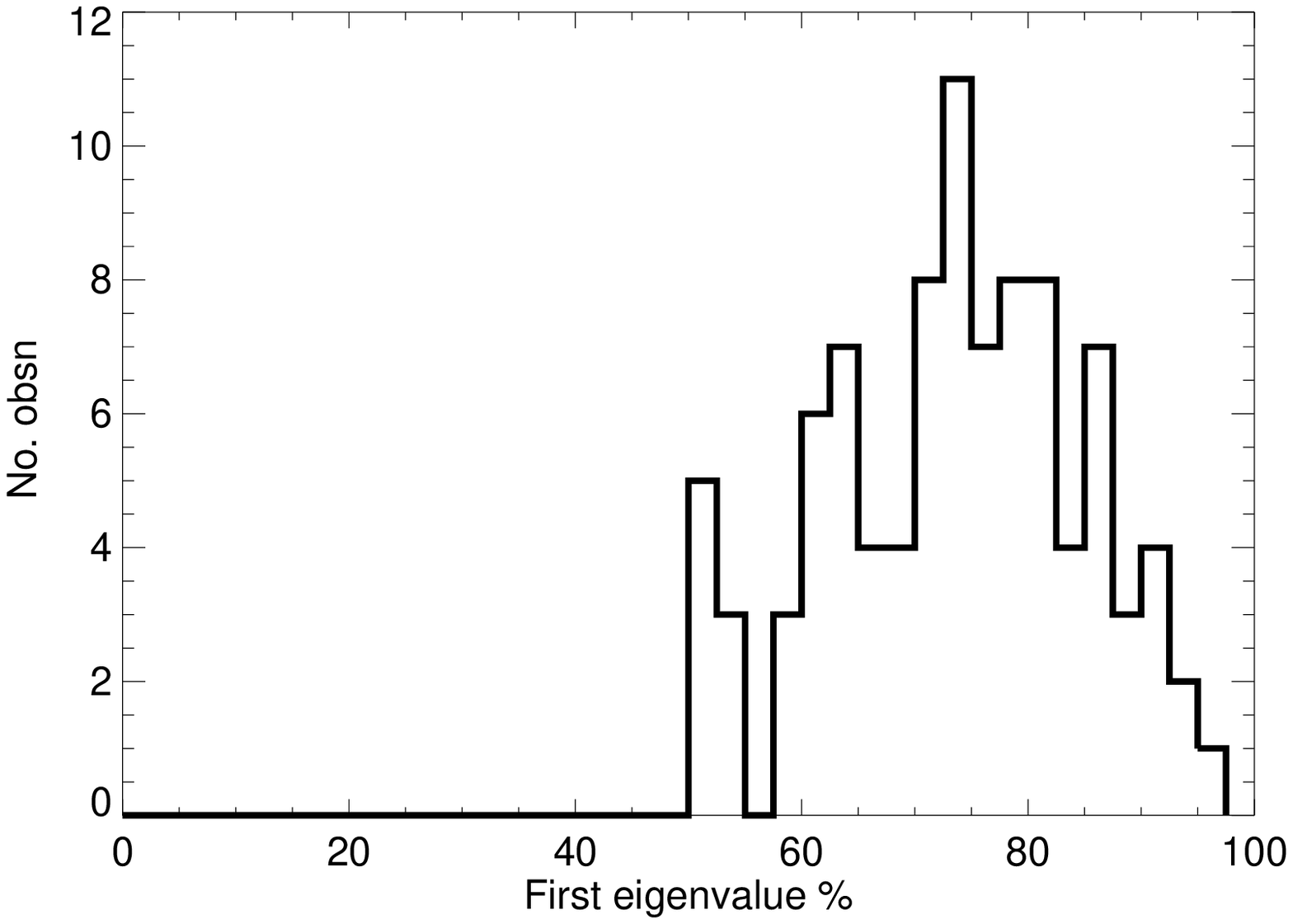}
  \caption{Histograms of the first eigenvalues percentage contribution
    to the total, for the modelled emission versus the \xmm\ line-band
    lightcurve (left) and versus solar wind flux (right).}
  \label{figpcas}
\end{figure*}

\begin{figure}
  \centering
  \includegraphics[width=.475\textwidth, angle=0, bb=70 360 530 695]{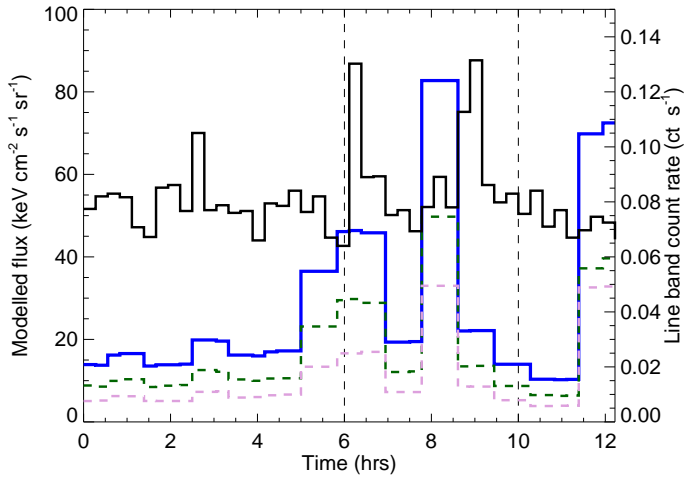}
  \caption{Example modelled lightcurve (blue, left-hand y-axis) with
    the \xmm\ line-band (black, right-hand y-axis), for the case where
    the first eigenvalue percentage contribution was the lowest when
    comparing the modelled flux and \xmm\ lightcurves. The
    contribution to the modelled lightcurve from the magnetosheath
    (green-dashed) and region past the bow shock (plum-dashed) are
    also shown. The SWCX-affected period was taken between the
    vertical dashed lines.}
  \label{figmostms}
\end{figure}

We wished to test how well the individual modelled flux lightcurve
tracked that of the line-band lightcurve for each observation. We also
wanted to determine the most dominant parameter in the modelling of
the expected emission. To do this we applied principal component
analysis to the model versus the line-band lightcurve and the model
versus the solar wind flux. We calculate the correlation matrix
between a linear fit to the relationship between each pair of values,
for each exospheric-SWCX case. We calculate the correlation rather
than the covariance matrix as the scale ranges of the data differ by a
large amount and so by using the correlation coefficients we
standardise the data. We use the primary eigenvalue of this matrix to
calculate the percentage contribution along the assumed linear
relationship between these lightcurves. Histograms of these percentage
contributions can be seen in Figure~\ref{figpcas}. The histogram
(left) shows that the X-ray lightcurve is generally correlated with
the modelled lightcurve, as the first principal component percentages
are high (with a mean of 73.7\,\%). The histogram (right) also shows
high first principal component percentages (with a mean of 73.6\,\%),
which suggests that in the vast majority of cases the model is
dominated by the incoming solar wind flux. The lowest eigenvalue when
comparing the modelled emission to the line-band lightcurve occurred
for the observation with the identification number
0101440401. Modelled and line-band lightcurves for this case are shown
in Figure~\ref{figmostms}. We also plot the component lightcurves that
make up the total modelled lightcurve from within the magnetosheath
and from beyond the bow shock. The contributions from the
magnetosheath region in this case dominates the modelled
lightcurve. \xmm\ is found anti-sunward of Earth during this
observation and so the line-of-sight of \xmm\ passed through the
flanks of the magnetosheath. This region is less well defined in our
model due to the approximations of the shape of the bowshock boundary
and the extrapolation of the values used to perturb both the solar
wind density and velocity in this region. The overall modelled
emission was very low for this case, compared with the top-row cases
of Figure~\ref{figexamplemod}.

\begin{figure}
  \centering
  \includegraphics[width=0.475\textwidth, angle=0, bb=77 370 553 700]{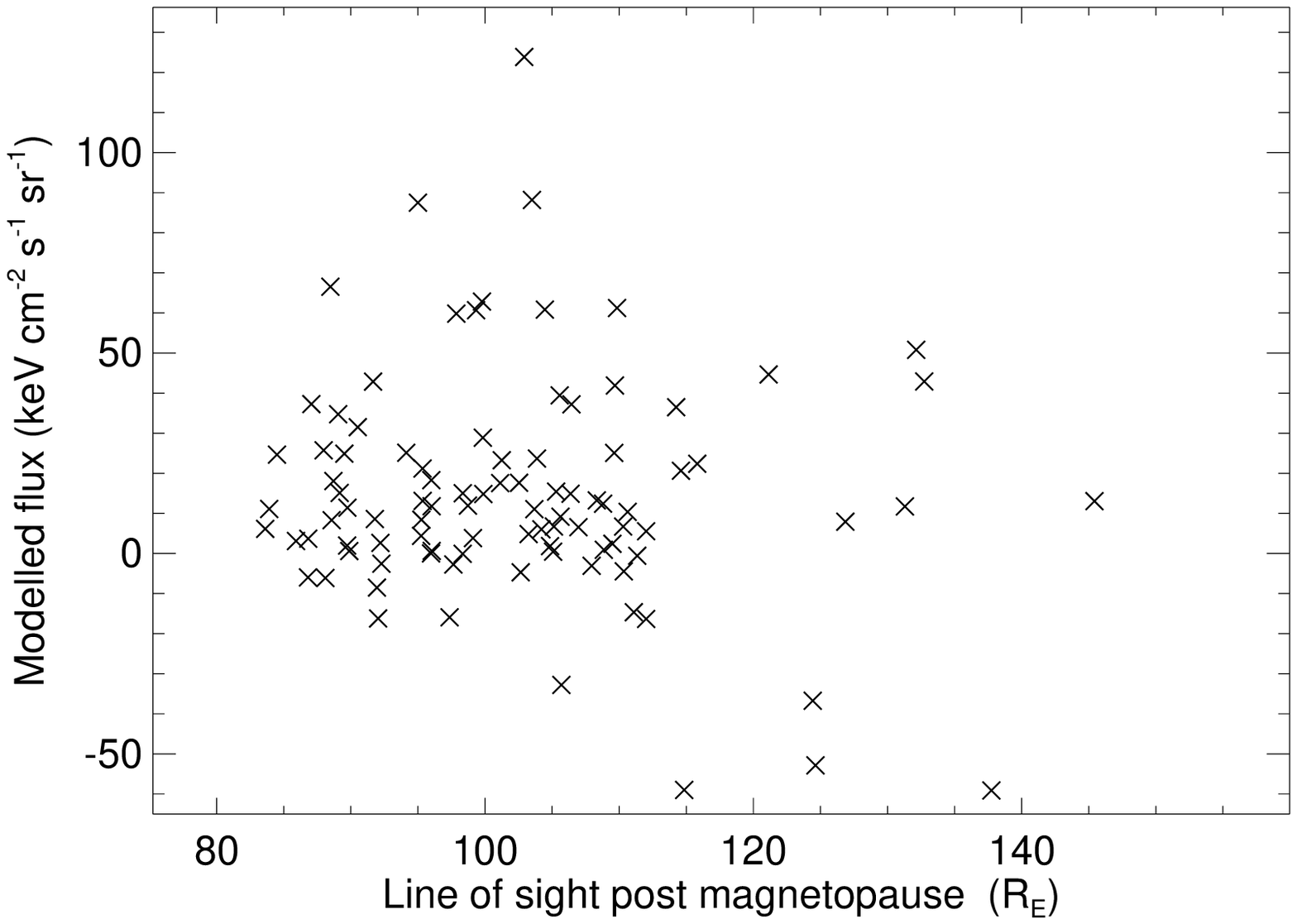}
  \includegraphics[width=0.475\textwidth, angle=0, bb=77 370 553 700]{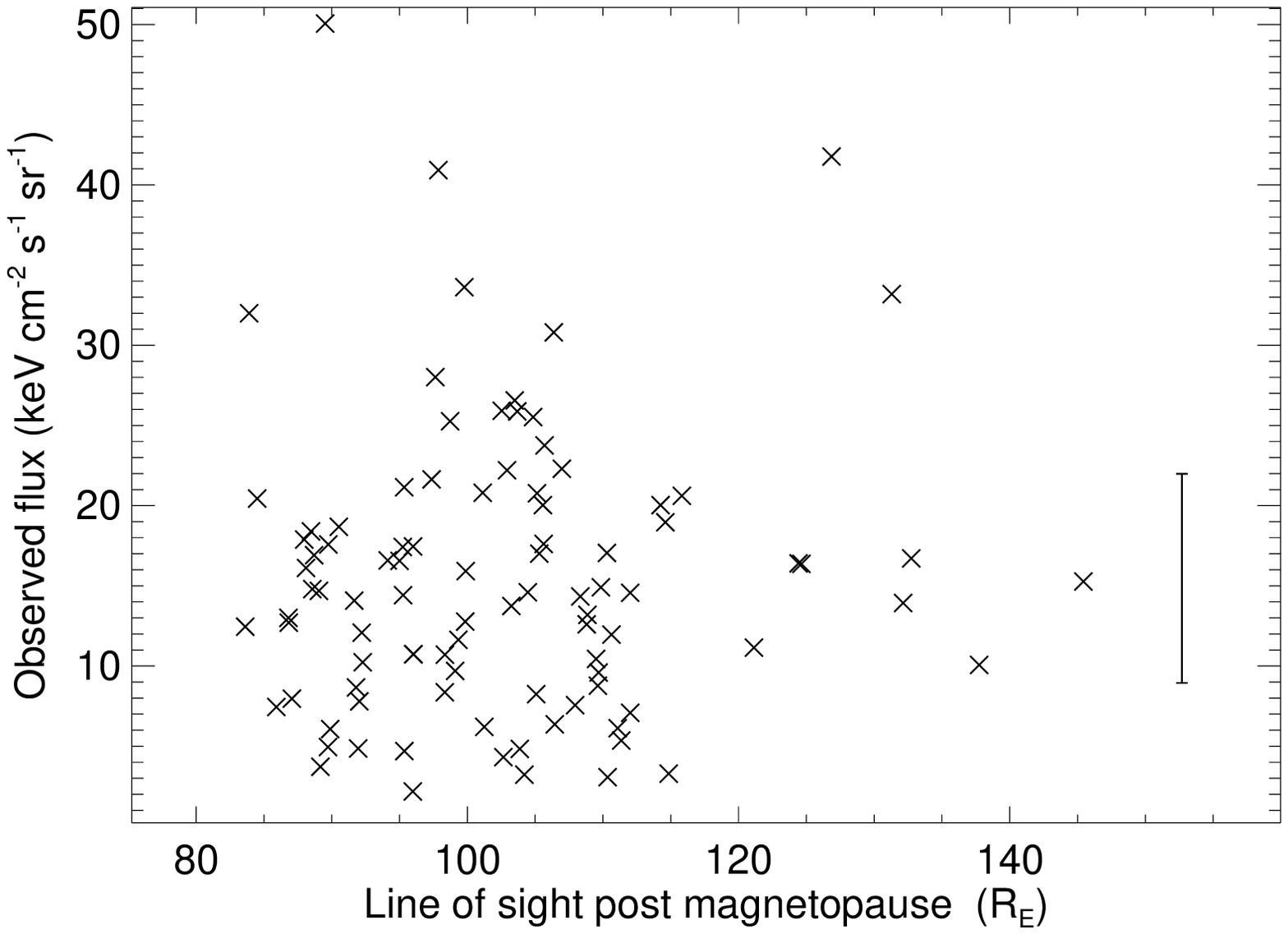}
  \caption{Line of sight length through the magnetosheath versus the
    modelled flux for the SWCX set (top) and the observed flux for the
    SWCX set (bottom). A mean error on the observed flux bar is given
    to the right of the bottom plot.}
  \label{figloslm}
\end{figure}

In Figure~\ref{figloslm} (top) we plot the modelled resultant flux
versus the average length of the line-of-sight through the
magnetosheath (between the magnetopause and the bow shock). In
Figure~\ref{figloslm} (bottom) we present the observed flux versus the
average length of the line-of-sight through the magnetosheath. There
is no discernible general relationship between the modelled or the
observed flux with the length of line-of-sight through the
magnetosheath.

\begin{figure}
  \centering
  \includegraphics[width=0.475\textwidth, angle=0, bb=60 360 535 616]{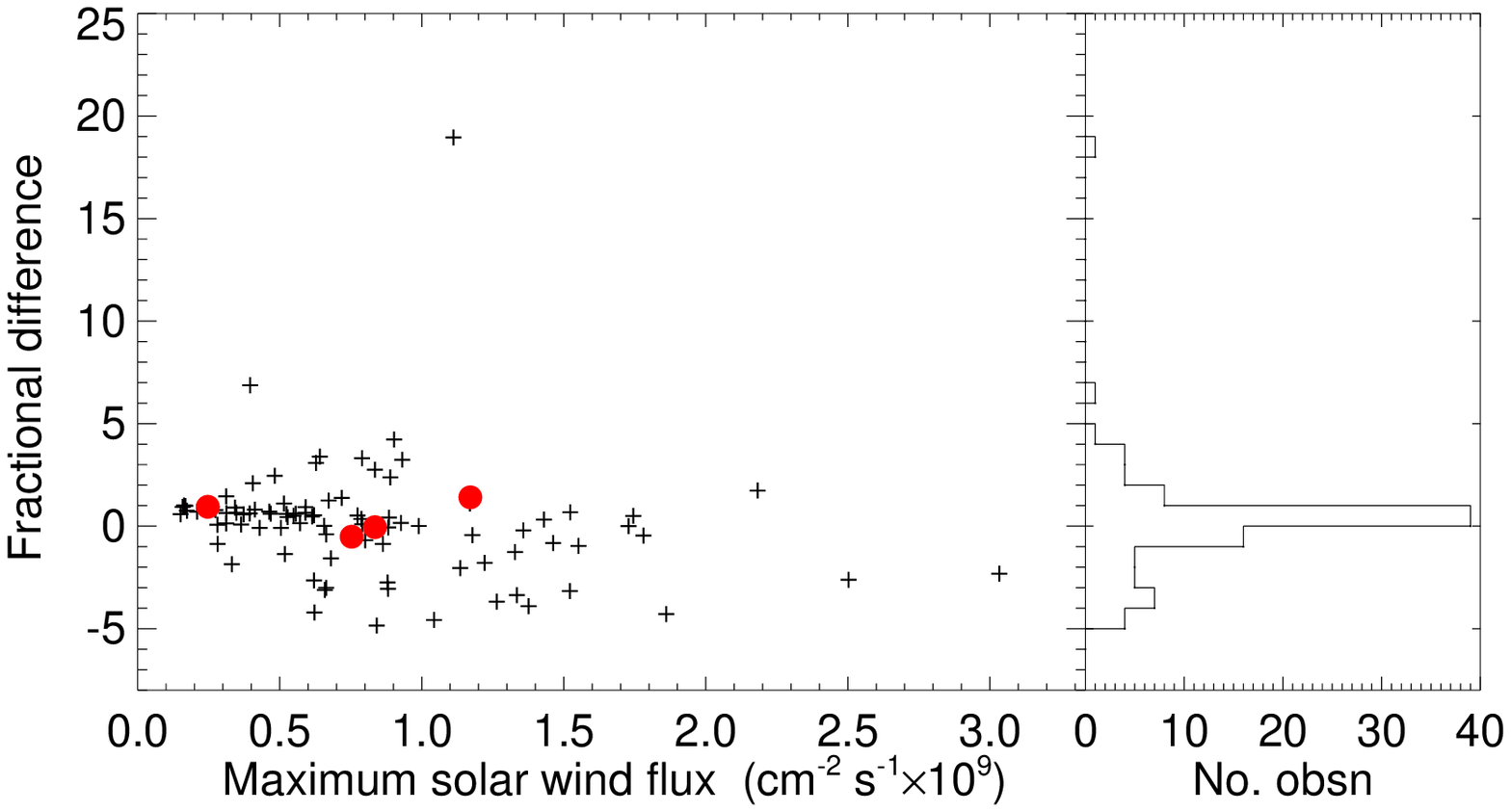}
  \includegraphics[width=0.475\textwidth, angle=0, bb=60 360 535 616]{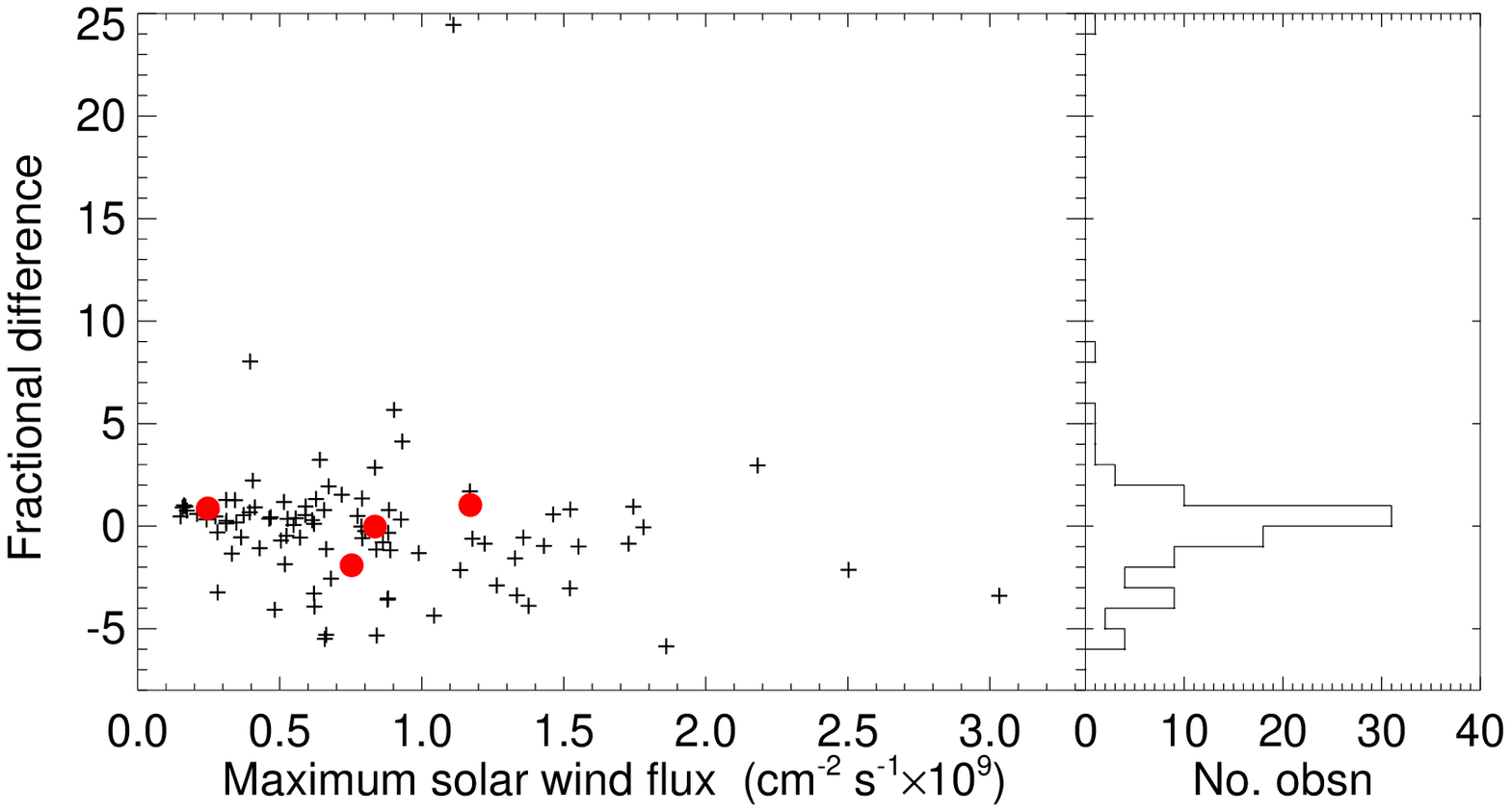}
  \caption{Fractional difference between (top) the observed and
    modelled flux and (bottom) the observed and \modtwo\ flux, versus
    the maximum solar wind flux. Also included in each panel
    is a histogram of the fractional differences. Cases where \xmm\ is
    found within the helium focusing cone are marked in red.}
  \label{figratioswmax}
\end{figure}

We investigate when the model and observed fluxes are discrepant by
calculating the fractional difference between the observed and
modelled fluxes ((observed-modelled)/observed flux). In
Figure~\ref{figratioswmax} (top panel) we plot this fractional difference versus
the maximum solar wind flux during each observation, along with a
histogram of the fractional difference values. The mean of these
fractional differences was +0.17 and the modal bin of the histogram
was for values between 0 and 1. A large proportion (approximately
60\,\%) of the modelled cases had a fractional difference between -1 and
1. The most discrepant cases occurred when the solar wind flux was low
(compared to the maximum solar wind flux of these exospheric-SWCX
cases). The solar wind plasma flow around the Earth's magnetosheath in
these cases has been badly described by the model. 

The observation with the largest absolute fractional difference had
identifier 0041750101. This case was similar to cases (bottom-left and
bottom-right) of Figure~\ref{figexamplemod} when the modelled
lightcurve peaked in the alternative (SWCX-free) period to the
enhancement in the observed line-band lightcurve (SWCX-affected).

We also wished to consider whether the fractional difference was due
to some underlying emission with temporal variability occurring in
near-heliospheric space, in particular to that of the helium focusing
cone \citep{weller1974}. We consider cases within the SWCX set that
occur within $10^{\circ}$ of the cone's direction ($73.9^{\circ}$
ecliptic longitude and $-5.6^{\circ}$ ecliptic latitude
\citep{witte1996}). As the integration length for the model is
relatively short compared to the spatial extent of the helium focusing
cone and size of Earth's orbit, only those observations taken when
\xmm\ is within this region are of importance. We find 4 cases within
this region. These cases are marked in red on
Figure~\ref{figratioswmax}. A statistical analysis, repeatedly drawing
4 random cases from the SWCX set, indicates that we obtain an average
fractional difference for the 4 random cases to be greater than that
of the 4 helium focusing cone cases 28\,\% of the time. We therefore
have no evidence to suggest that temporal variability originating in
the helium focusing cone is a significant component of the
observed-to-modelled flux discrepancy.

We also compute the fractional differences between the observed and
modelled flux values for the \modtwo\ results. These are shown in
Figure~\ref{figratioswmax} (bottom panel). The peak of the
distribution lies in the same bin as that of
Figure~\ref{figratioswmax} (top panel), although there is a greater
variance seen in the differences. We conclude that for the SWCX set
cases, no benefit has arisen by using a compositionally-variable
dependent model as opposed to the simple model. We continue our
discussion based on the simple model results only.

\begin{figure*}
  \centering
  \includegraphics[width=0.33\textwidth, angle=0, bb=90 370 543 700]{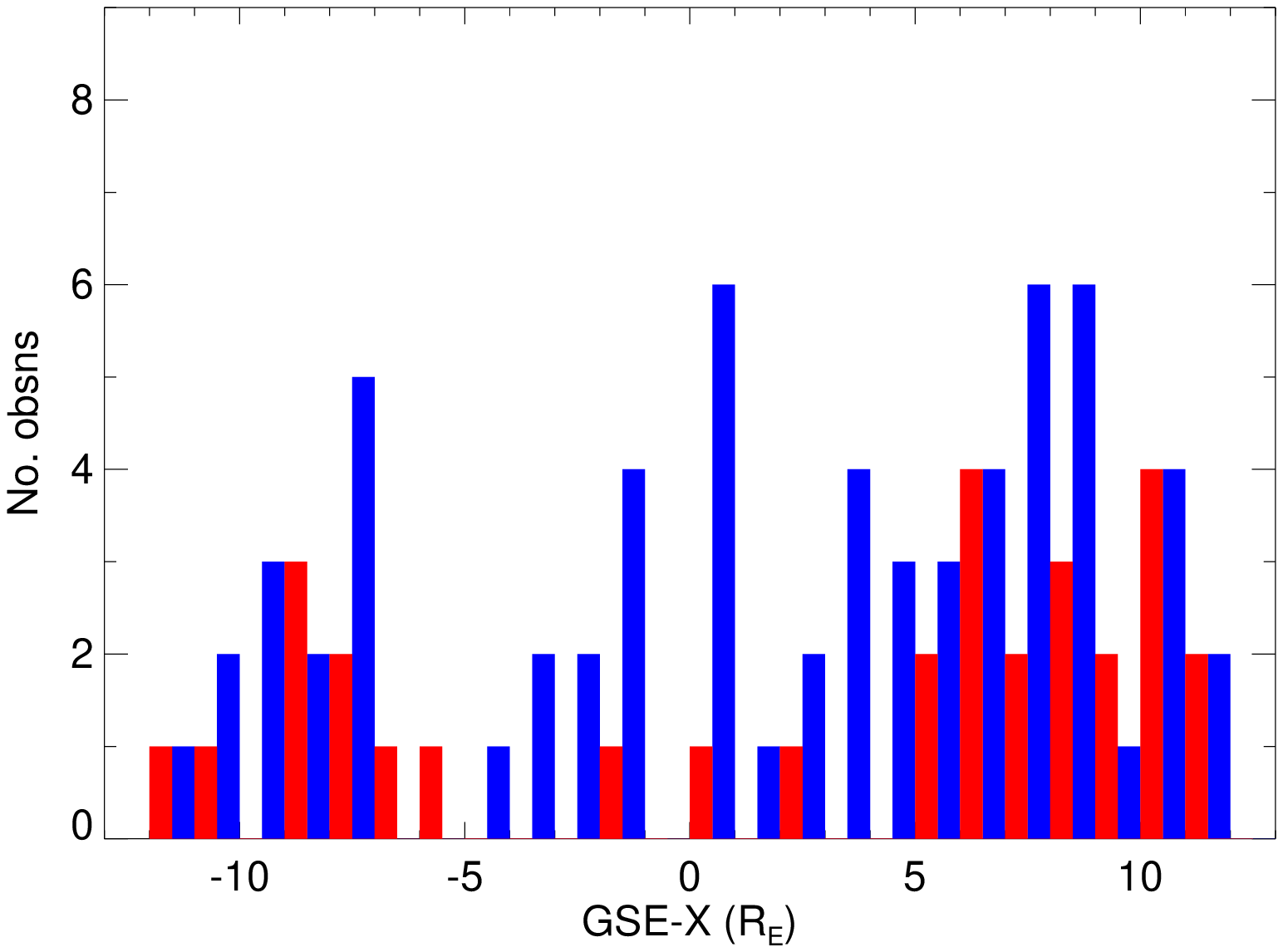}
  \includegraphics[width=0.33\textwidth, angle=0, bb=90 370 543 700]{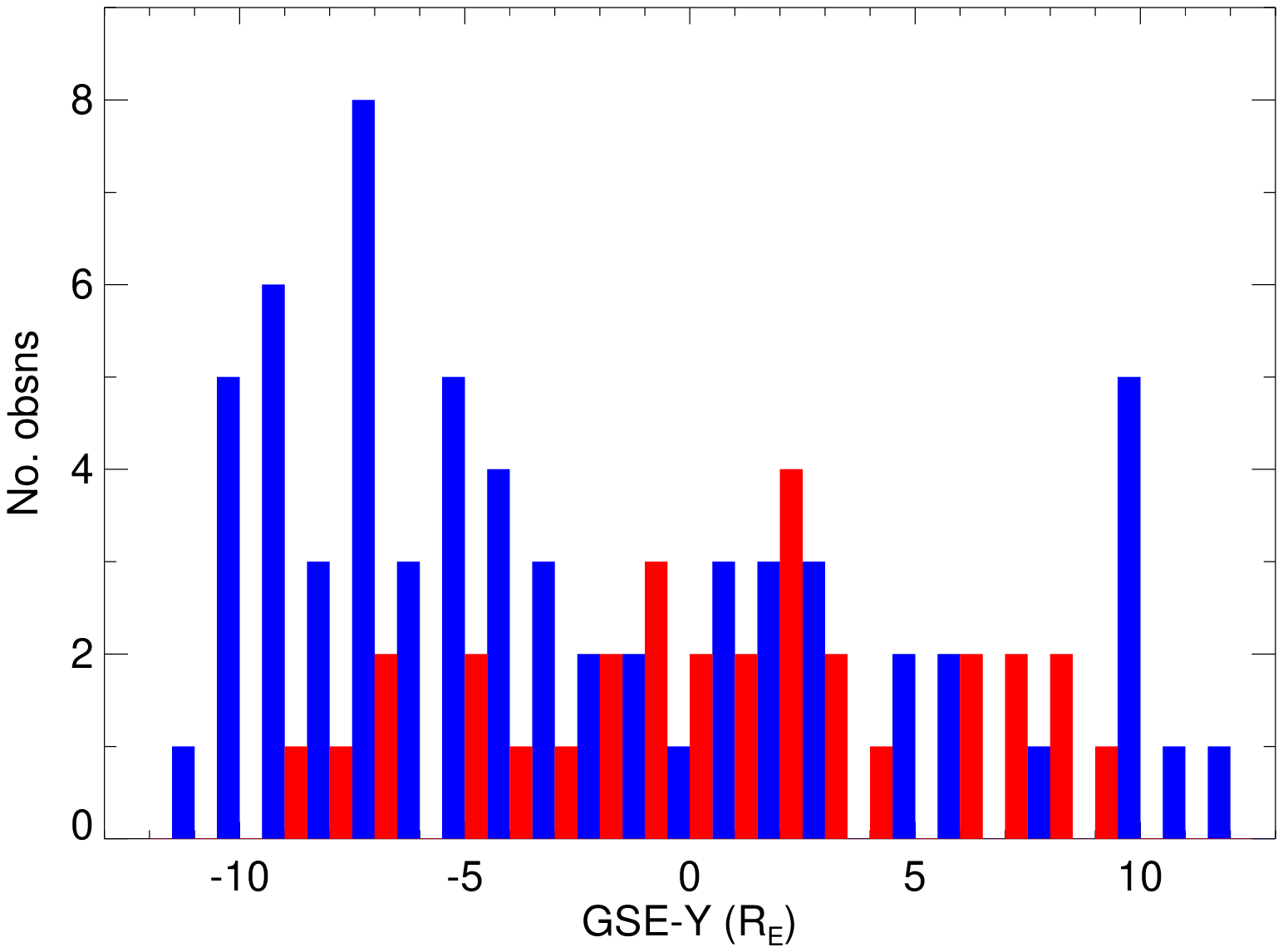}
  \includegraphics[width=0.33\textwidth, angle=0, bb=90 370 543 700]{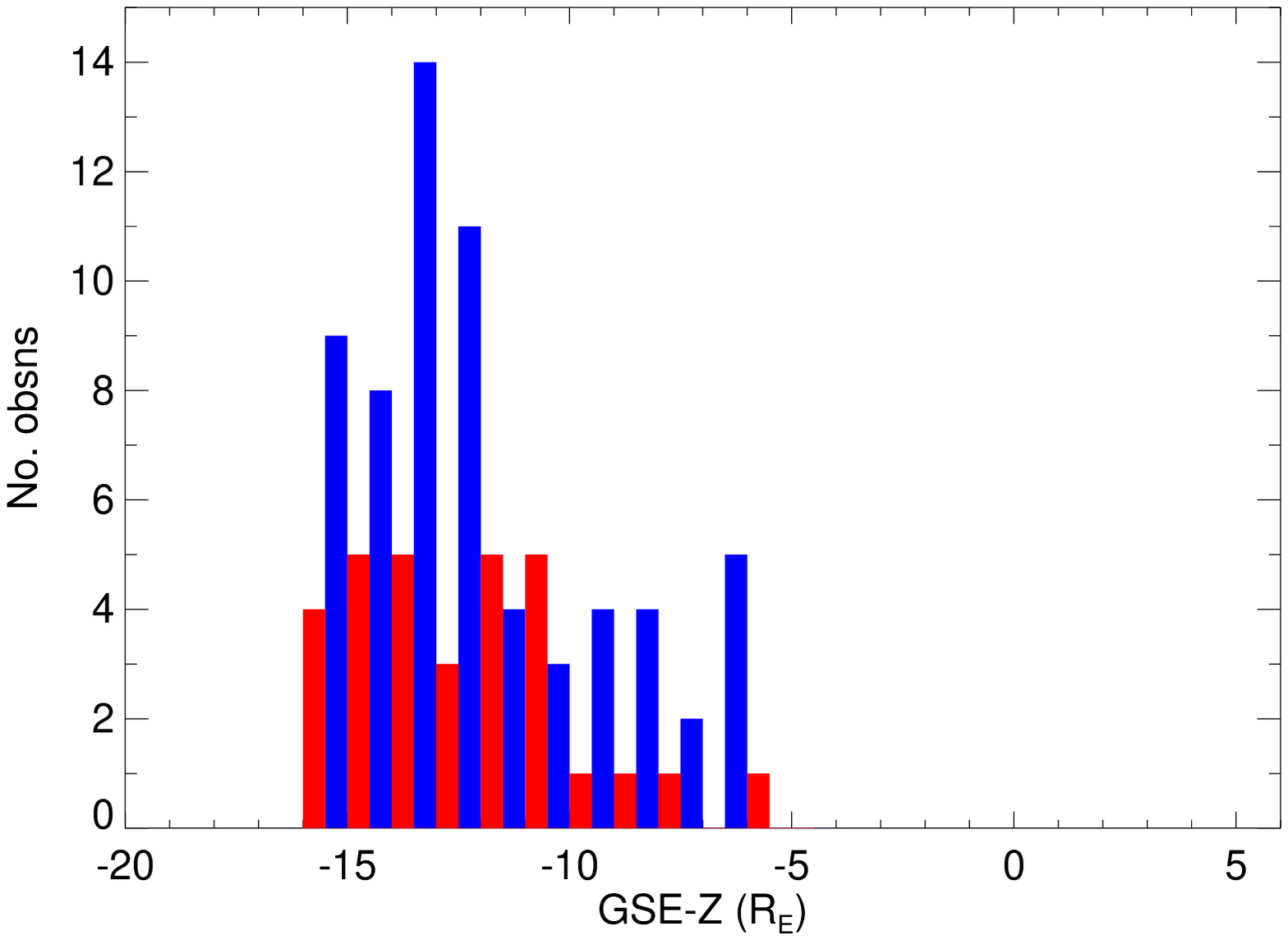}
  \caption{Histograms of mid-observation (left) GSE-X, (middle) GSE-Y
    and (right) GSE-Z \xmm\ positions for good (blue) and bad (red)
    fractional differences between the observed and modelled
    fluxes. The histogram bins have been offset from one another in
    the plot, for ease of viewing.}
  \label{figfracvsgsey}
\end{figure*}

We split the fractional difference values into two sets; for cases
where this value is $<$\,-1.5 or $>$\,1.5 (bad), or any other value
(good). In Figure~\ref{figfracvsgsey} we plot histograms of the
mid-observation position of \xmm\ (in GSE coordinates, GSE$\mhyphen$X,
Y and Z) for each observation for the good and bad sets. We performed
a Kolmogorov-Smirnov test on the three pairs of good and bad sets. The
probabilities that the good and bad sets are drawn from the same
sample distribution were 0.22, 0.002 and 0.79 for GSE$\mhyphen$X,
GSE$\mhyphen$Y and GSE$\mhyphen$Z respectively, indicating that for
the GSE$\mhyphen$Y coordinates, the good and bad sets are
statistically different. The good set for the GSE$\mhyphen$Y positions
are skewed towards negative values and there are relatively more
observations in the bad set in the positive direction. We repeat the
test using mid-observation position of \xmm\ expressed in Geocentric
Solar Magnetospheric (GSM) coordinates. These differ from the GSE
coordinates as the GSM$\mhyphen$Y axis is perpendicular to the Earth's
magnetic dipole (the X$\mhyphen$axis is unchanged). The
Kolmogorov-Smirnov test results were 0.22, 0.004 and 0.28 for the
GSM$\mhyphen$X, GSM$\mhyphen$Y and GSM$\mhyphen$Z respectively. The
Y$\mhyphen$coordinate result remains significant. Therefore we
postulate that the model is better at describing the conditions seen
by \xmm\ when the Y$\mhyphen$coordinate is negative.

The simplifications used in this model to describe the flanks of the
magnetosheath in terms of shape, solar wind density and velocity may
mean that the model is less robust in this region. We assumed
cylindrical symmetry about the GSE$\mhyphen$X axis, however, the
magnetosheath will be non-symmetrical in shape, suffering for example
magnetosheath erosion along one side of the magnetopause (along the
dusk side, or GSE$\mhyphen$Y \citep{owen2008}), a full discussion of
which is beyond the scope of this paper. The incoming solar wind is
expected from the GSE$\mhyphen$Y positive direction, determined by the
flow of the solar wind along the Parker Spiral as it emanates from the
Sun. It is in this region that we expect the greatest differences in
shape from the simplified magnetosheath we have used in our modelling
steps and it is here that we see the largest absolute fractional
differences between the observed and modelled fluxes. It is clear that
although the model can estimate the observed flux within a factor of
$\sim$\,2 in approximately 50\,\% cases, there are still many
occurrences when the local physical conditions combine so that the
simple model does not explain the observed flux adequately.

\section{Discussions and conclusions}\label{secconcl}

We have identified \nthisswcx\ \xmm\ observations, \pcswcx\,$\%$ of
the sample studied, when temporally variable SWCX emission was present
in the data. The method of this paper has been able to identify cases
of temporally variable SWCX from within a large sample of \xmm\
observations. These cases were taken from those observations
presenting the highest \redc\ and \cratio\ values. The corresponding
occurrence rate within the sample used in Paper I was
$\sim$\pcoldswcx\,\%. The data for this paper covered a wider range in
time compared to \papone. The level of detection can be attributed to
the reduction in solar activity as this time range extended into a
period towards solar minimum. There will be many more \xmm\
observations affected by SWCX, either occurring within the exosphere
or near-interplanetary space, such as within the helium focusing cone,
or at the heliospheric boundary and undetectable here. SWCX occurring
within the heliosheath will generally vary over longer periods than
exospheric SWCX and so is more suited to detection by
observation-to-observation comparison (e.g. observations within the
studies of \citet{kuntz2008} and \citet{henley2010}). Enhancements
from the helium focusing cone will produce some temporal variation but
are strongly constrained by viewing geometry. The method presented in
this paper is only able to identify time-variable SWCX which varies
over the length of an observation and therefore the level of
contamination quoted in this paper can only provide a lower limit to
the occurrence of exospheric SWCX as observed by \xmm. When SWCX
emission is only slowly varying or constant over an exposure it will
be undetectable by this method. There will also be cases which have
slipped detection due to a high percentage of the observation data
being removed by the flare-filtering process, resulting in short
lightcurves that are excluded from our analysis. As increased
concentrations of solar wind ions in the magnetosheath are expected to
mirror increases in the general flux of the solar wind, increased
levels of SWCX emission are expected precisely when the flux of solar
proton increases. If the on-board radiation monitors of \xmm\ detect a
dangerous environment for the satellite, the science instruments are
switched into a safe mode which invariably leads to the loss of high
SWCX emission periods being available for detection within our
sample. Even after soft-proton flare-filtering has been applied to the
data, considerable proton-contamination may be present. This can
result in a significant scatter when plotting either the line-band or
continuum lightcurve, whilst potentially masking a clear enhanced
period of SWCX-emission during the observation. The level of residual
soft-proton contamination may mean that the observation is completely
rejected by an observer. If the user does indeed proceed to process
the data, the limits presented here on \redc\ and \cratio\ may be
useful to guide any further analysis as to whether extra caution
should be taken to account for potentially high levels of
time-variable SWCX contamination.

We have shown that exospheric-SWCX occurs preferentially on the
sunward side of the magnetosheath, when the line-of-sight of the \xmm\
pointing towards its astronomical target of interest intersected the
area of strongest expected X-ray emission of the exosphere. This
occurs during the northern hemisphere summer months. However, a
considerable fraction of the SWCX-affected observations had
lines-of-sight that intersected the flanks of the magnetosheath, where
the SWCX X-ray emission is expected to be weaker. The example
presented in \citet{carter2010}, along with showing the highest flux
of the SWCX set, is one such case whereby \xmm\ was not pointing in
the region of strongest expected X-ray flux. This suggests that there
are considerable deviations from our current understanding of either
or both the hydrogen neutral density and the perturbation of the solar
wind in the flank regions of the magnetosheath. A dedicated mission
observing SWCX emission to probe the magnetosheath would answer many
questions regarding the distribution of mass and mass transfer in the
magnetosheath, bowshock and near vicinity of the Earth
\citep{collier2010}.

For each time-variable exospheric-SWCX case and EPIC-MOS instrument,
spectra were created for the SWCX-affected and the SWCX-free
periods. The resulting difference spectrum between the two periods
became the spectrum used for further spectral analysis. We applied, to
each difference spectrum, a standardised spectral model of 33 Gaussian
lines involving 9 ion species. We set the relative normalisations
between lines for transitions for one particular species to the ratios
of laboratory cross-sections measured for a collisional speed of
400\,$\mathrm{km}\,\mathrm{s}^{-1}$ between ions and atomic hydrogen. A combined
EPIC-MOS flux was calculated between 0.25 and 2.5\,keV for each
case. The SWCX set showed a large spread in spectrally modelled
observed flux. Although the mean solar proton flux during the
SWCX-affected period was not a very good indicator of the level of
observed flux, there was a positive correlation between these two
parameters.

The SWCX set showed a range of spectral characteristics, with \oxys\
and \oxye\ being the dominant lines. Spectral signatures obtained from
these \xmm\ observations, such as the ratio between magnesium and
oxygen ion species, may be complementary to data obtained from in-situ
solar wind monitors in classifying solar wind plasma types. We have
studied the SWCX set with the largest observed flux in a separate
paper \citep{carter2010} which we attributed to a CME passing by the
Earth. This case also formed part of the subset that exhibited the
highest \mage\ to \oxys\ and \oxye\ to \oxys\ ratios. CME plasma is
compositionally different to steady state solar wind plasma. Other
phenomena, such as co-rotating interacting regions for example, may
include high density pulses of plasma but show spectral signatures
close to canonical solar wind plasma conditions.

We wished to investigate whether the observed spectrally modelled flux
could be estimated using a simple model, constructed using data
describing upwind solar wind conditions, and the orbital and target
pointing configuration of \xmm\ at the time of each SWCX-affected
observation. We used simple models of hydrogen densities about the
Earth and the perturbations of the solar wind within the region of the
magnetosheath. Approximately 60\,\% of exospheric-SWCX cases showed an
observed to modelled flux fractional difference between -1 and
1. Negative values of the modelled flux occurred when the model
predicted an emission pulse in the alternative time period to that
assigned as the SWCX-affected period. The largest outliers occurred
when the solar wind flux was at its weakest. The model was dominated
by the solar wind flux. The presence of the magnetosheath made a large
contribution to the modelled emission in a few cases. The actual
line-of-sight length through the magnetosheath did not have any
discernable influence on either the observed or modelled flux. The
model employed a large parameter space and there are various aspects
which are expected to have a large uncertainty (such as the
calculation of the delay from \ace\ to the Earth, if the solar wind
plasma front is tilted or the distribution of solar wind flow around
the magnetosheath, especially in the regions far from the subsolar
point). Adapting the model to account for changes in the solar wind
\opluss/\oplussix\ ratio did not improve the observed to modelled flux
fractional difference for the SWCX set overall. We have not accounted
for any anisotropies in the Earth's exosphere in terms of hydrogen
density. In addition there was some suggestion that those cases when
\xmm\ was found at positive GSE$\mhyphen$Y (the dusk side) resulted in
the least well-fitting models, where anisotropies in the shape of the
magnetosheath may be most apparent.

The authors, who are members of the \xmm\ EPIC BGWG, intend to provide
a list detailing the observations affected by time-variable SWCX in
the near future, as part of the BGWG web pages and group activities.

\section{Acknowledgements}\label{secackno}
We thank Anthony Williams for helpful discussions regarding
constituents of the solar wind. We thank the anonymous referee for
their suggestions that have greatly enhanced this paper. The authors
gratefully acknowledge funding by the Science and Technology
Facilities Council, U.K.

\bibliographystyle{aa} 
\bibliography{phd_gen_aa} 

\appendix
\section{List of exospheric-SWCX affected XMM-{\it Newton}
  observations}\label{secappendix}
\begin{table*}
  \caption[]{Table of the SWCX set observations, ranked by \redc\ (the reduced-$\chi^{2}$ to the linear fit between the line-band and continuum lightcurves). Also listed for each case are the revolution number (Revn), observation (Obsn) and the MOS1 and MOS2 exposure identifiers (Expn M1 and Expn M2) and the ratio of the lightcurve variances (\cratio).}
    \begin{tabular}{lllllll|lllllll}
\hline\hline
Ind. & Revn & Obsn & Expn M1 & Expn M2 & \redc\ & \cratio\  & Ind. & Revn & Obsn & Expn M1 & Expn M2 & \redc\ & \cratio\ \\
\hline
1 & 0342 & 0085150301 & U003 & U003 & 27.2 & 10.3 &  53 & 0997 & 0206090201 & S001 & S002 & 2.5 & 2.6 \\
2 & 0209 & 0093552701 & S001 & S002 & 23.0 & 4.0 &  54 & 0690 & 0134531701 & S002 & S003 & 2.4 & 2.4 \\
3 & 1014 & 0305920601 & S001 & S002 & 15.0 & 30.9 &  55 & 1023 & 0304531801 & S001 & S002 & 2.4 & 2.2 \\
4 & 0690 & 0149630301 & S001 & S002 & 14.1 & 21.6 &  56 & 0476 & 0109661201 & S001 & S002 & 2.4 & 2.6 \\
5 & 0623 & 0150610101 & U002 & U002 & 13.5 & 4.8 &  57 & 1018 & 0212480801 & S001 & S002 & 2.4 & 8.8 \\
6 & 0339 & 0054540501 & S002 & S003 & 13.2 & 22.4 &  58 & 0871 & 0206360101 & S001 & S002 & 2.4 & 1.3 \\
7 & 0422 & 0113050401 & S001 & S002 & 12.7 & 12.3 &  59 & 0689 & 0149610401 & S001 & S006 & 2.4 & 1.9 \\
8 & 0151 & 0094800201 & S001 & S002 & 12.6 & 7.2 &  60 & 0859 & 0203541101 & S001 & S002 & 2.3 & 2.0 \\
9 & 0657 & 0141980201 & S001 & S002 & 12.0 & 8.1 &  61 & 0354 & 0049340201 & S001 & S002 & 2.3 & 1.8 \\
10 & 0664 & 0150680101 & S001 & S002 & 9.8 & 5.3 &  62 & 0052 & 0099760201 & S001 & S002 & 2.2 & 1.0 \\
11 & 0505 & 0153752201 & S002 & S003 & 8.5 & 6.5 &  63 & 0322 & 0094400101 & S001 & S002 & 2.2 & 2.3 \\
12 & 0271 & 0111550401 & S002 & S005 & 7.8 & 6.9 &  64 & 1232 & 0406420401 & S001 & S002 & 2.1 & 2.3 \\
13 & 0279 & 0070340501 & S001 & S002 & 7.8 & 2.3 &  65 & 0990 & 0203450201 & S001 & S002 & 2.1 & 4.7 \\
14 & 0178 & 0101040301 & S001 & S002 & 7.2 & 5.2 &  66 & 0167 & 0106460101 & S001 & S002 & 2.1 & 1.1 \\
15 & 0139 & 0109060101 & S002 & S003 & 7.0 & 6.1 &  67 & 0395 & 0084140501 & S002 & S003 & 2.1 & 5.8 \\
16 & 0529 & 0147540101 & S001 & S002 & 6.9 & 7.1 &  68 & 0234 & 0069750101 & S001 & S002 & 2.1 & 1.7 \\
17 & 1199 & 0402250201 & S001 & S002 & 6.8 & 9.5 &  69 & 0175 & 0110660401 & S002 & S003 & 2.1 & 2.0 \\
18 & 0676 & 0049540401 & S001 & S002 & 6.5 & 8.9 &  70 & 0554 & 0056021001 & S001 & S002 & 2.0 & 3.5 \\
19 & 0982 & 0306700301 & S001 & S002 & 6.2 & 9.9 &  71 & 1232 & 0405210601 & S001 & S002 & 2.0 & 1.5 \\
20 & 0645 & 0150320201 & S001 & S002 & 5.8 & 4.2 &  72 & 0150 & 0105260501 & S001 & S002 & 2.0 & 3.0 \\
21 & 0630 & 0143150601 & U002 & U002 & 5.7 & 8.0 &  73 & 0634 & 0151400201 & S001 & S002 & 2.0 & 3.4 \\
22 & 0494 & 0109120101 & S002 & S003 & 5.4 & 7.8 &  74 & 1594 & 0560191501 & S001 & S003 & 2.0 & 1.9 \\
23 & 0178 & 0110980101 & S001 & S002 & 5.2 & 1.9 &  75 & 0747 & 0200730401 & S001 & S002 & 1.9 & 1.1 \\
24 & 0114 & 0127921101 & S001 & S002 & 4.9 & 3.8 &  76 & 0428 & 0112520101 & S001 & S002 & 1.9 & 1.2 \\
25 & 0811 & 0202100301 & S001 & S002 & 4.5 & 3.9 &  77 & 0692 & 0112490301 & S011 & S012 & 1.8 & 1.5 \\
26 & 0997 & 0303260501 & S001 & S002 & 4.1 & 2.3 &  78 & 0428 & 0112521001 & S001 & S002 & 1.8 & 1.5 \\
27 & 0163 & 0100640201 & S002 & S003 & 3.7 & 3.7 &  79 & 0875 & 0203750101 & S001 & S002 & 1.8 & 1.9 \\
28 & 0431 & 0136000101 & S002 & S003 & 3.6 & 2.2 &  80 & 0457 & 0124712501 & S002 & S003 & 1.8 & 2.9 \\
29 & 0605 & 0146390201 & S001 & S002 & 3.5 & 4.8 &  81 & 0191 & 0093550401 & S001 & S002 & 1.8 & 1.8 \\
30 & 0906 & 0203361501 & S001 & S002 & 3.4 & 2.7 &  82 & 0882 & 0203610401 & S001 & S003 & 1.8 & 2.2 \\
31 & 0113 & 0127921001 & S001 & S002 & 3.4 & 2.0 &  83 & 0865 & 0206610201 & S001 & S002 & 1.7 & 1.2 \\
32 & 0834 & 0200000101 & S001 & S002 & 3.3 & 3.4 &  84 & 1349 & 0406960101 & S001 & S002 & 1.7 & 1.1 \\
33 & 0846 & 0164560701 & S001 & S002 & 3.3 & 1.8 &  85 & 0750 & 0201160401 & S001 & S002 & 1.6 & 1.8 \\
34 & 0387 & 0073140501 & S004 & S005 & 3.2 & 1.3 &  86 & 0235 & 0051940501 & S001 & S002 & 1.6 & 2.6 \\
35 & 1600 & 0553650101 & S001 & S002 & 3.1 & 2.3 &  87 & 0750 & 0201030301 & S001 & S002 & 1.6 & 3.4 \\
36 & 0555 & 0146510301 & S001 & S002 & 3.0 & 3.5 &  88 & 0420 & 0093190501 & S001 & S002 & 1.5 & 1.0 \\
37 & 0515 & 0089370501 & S001 & S002 & 2.9 & 2.3 &  89 & 0457 & 0112521301 & S001 & S002 & 1.5 & 2.5 \\
38 & 1049 & 0300800101 & S002 & S003 & 2.9 & 1.4 &  90 & 0168 & 0101440401 & S001 & S002 & 1.5 & 1.6 \\
39 & 0376 & 0001930301 & S001 & S002 & 2.9 & 1.6 &  91 & 0369 & 0084230201 & S001 & S002 & 1.5 & 1.3 \\
40 & 1206 & 0404965401 & S003 & S004 & 2.9 & 1.6 &  92 & 0461 & 0041750101 & S001 & S003 & 1.5 & 1.7 \\
41 & 0574 & 0110910201 & S002 & S003 & 2.9 & 1.3 &  93 & 0997 & 0201330101 & S001 & S002 & 1.5 & 1.2 \\
42 & 0136 & 0101440101 & S001 & S002 & 2.9 & 1.7 &  94 & 0630 & 0151390101 & S007 & S008 & 1.5 & 1.2 \\
43 & 1075 & 0305560101 & S001 & S002 & 2.8 & 2.7 &  95 & 0449 & 0082140301 & S001 & S002 & 1.5 & 1.8 \\
44 & 0643 & 0141150101 & S001 & S002 & 2.7 & 2.4 &  96 & 0325 & 0085280501 & S001 & S002 & 1.4 & 1.5 \\
45 & 0918 & 0206430101 & S001 & S002 & 2.7 & 2.2 &  97 & 1555 & 0552410401 & S001 & S002 & 1.4 & 1.0 \\
46 & 0313 & 0092140101 & S001 & S002 & 2.7 & 2.6 &  98 & 0148 & 0112880801 & S001 & S002 & 1.4 & 1.4 \\
47 & 0391 & 0085280301 & S001 & S002 & 2.7 & 1.8 &  99 & 0159 & 0112980201 & S001 & S002 & 1.4 & 1.3 \\
48 & 0173 & 0106660201 & S001 & S002 & 2.7 & 1.2 &  100 & 0676 & 0152460301 & S001 & S002 & 1.3 & 2.8 \\
49 & 0982 & 0303720301 & S001 & S002 & 2.6 & 2.8 &  101 & 0974 & 0302640101 & S002 & S003 & 1.3 & 1.8 \\
50 & 0258 & 0112290201 & S001 & S002 & 2.6 & 3.4 &  102 & 1094 & 0306680201 & S001 & S002 & 1.3 & 1.1 \\
51 & 0260 & 0070340201 & U002 & U002 & 2.6 & 3.3 &  103 & 1364 & 0500500801 & U002 & U002 & 1.3 & 1.7 \\
52 & 0484 & 0103060201 & S002 & S003 & 2.5 & 1.7 &  &  &  &  &  &  \\
    \end{tabular}
    \label{tabappendix}
\end{table*}

\end{document}